\documentclass[12pt]{article}
\usepackage{axodraw}
\usepackage{defsign,defref}
\usepackage{amsmath,amssymb}
\usepackage{graphicx}
\usepackage[nosort]{cite}
\newcommand{\preprint}[1]{\begin{flushright}#1\end{flushright}}

\newcommand \ket[1]{\left\vert\, {#1} \, \right>}

\newcommand{\bea}{\begin{eqnarray}}
\newcommand{\eea}{\end{eqnarray}}
\newcommand{\simgt}{\hbox{ \raise3pt\hbox to 0pt{$>$}\raise-3pt\hbox{$\sim$} }}
\newcommand{\simlt}{\hbox{ \raise3pt\hbox to 0pt{$<$}\raise-3pt\hbox{$\sim$} }}
\newcommand \vc[1]{{\bf {#1}}}

\def\to{\rightarrow}

\begin{document}
\preprint{TU-585\\Jan 2000}
\vspace*{3cm}
\begin{center}
  {\bf\large Probe of {\it CP} Violation in $e^+e^- \to t\bar{t}$
Near Threshold}
  \\[10mm]
  {M.~Je\.zabek 
    }
  \\[5mm]
  {\it \small
Institute of Nuclear Physics,
Kawiory 26a, PL-30055 Cracow, Poland
}
\\  {\small and} \\
{\it \small  
Department of Field Theory and Particle Physics, University of 
      Silesia, \\
     Uniwersytecka 4, PL-40007 Katowice, Poland
    }
  \\[5mm]
  {
    T. Nagano and Y. Sumino
    }
  \\[5mm]
  {\it \small
    Department of Physics, Tohoku University\\
    Sendai, 980-8578 Japan
    }
\end{center}
\vspace{1cm}
\begin{abstract}
We study how to probe the anomalous 
{\it CP}-violating couplings of top quark with $\gamma$,
$Z$ and $g$ in the $t\bar{t}$ threshold region at
future $e^+e^-$ colliders.
These couplings contribute to the difference
of the $t$ and $\bar{t}$ polarization vectors $\delta \vc{P}$
and to the {\it CP}-odd
spin correlation tensor $\delta\PolQBR_{ij}$.
We find that typical sizes of $\delta \vc{P}$ and 
$\delta\PolQBR_{ij}$ are 5--20\% times the couplings
$( \dtp , \dtZ , \dtg  )$ in the threshold region.
Experimentally $\delta \vc{P}$ can be measured efficiently using the
{\it CP}-odd combination of the
$\ell^\pm$ momenta or of the $\ell^\pm$ directions.
%We made rough estimates of sensitivities to the anomalous
%couplings expected at future $e^+e^-$ colliders:
%$\delta^{\rm (stat)} d_{tg}, \, \,
%\delta^{\rm (stat)} d_{t\gamma}, \, \,
%\delta^{\rm (stat)} d_{tZ}
%\sim \Order{10\%}$
%for an integrated luminosity of $50~{\rm fb}^{-1}$.
We have similar sensitivities to both the real and
imaginary parts of the couplings
independently using the two components
of $\delta \vc{P}$.
Taking advantage of 
different dependences of $\delta \vc{P}$ on
the $e^\pm$ polarizations and on the c.m.\ energy,
we will be able to disentangle the effects of the
three couplings $\dtp$, $\dtZ$, $\dtg$ in the
$t\bar{t}$ threshold region.  
We give rough estimates of sensitivities to the anomalous
couplings expected at future $e^+e^-$ colliders.
The sensitivities to $\dtp$ and $\dtZ$ are comparable
to those attainable in the open-top region at $e^+e^-$
colliders.
The sensitivity to $\dtg$ is worse than that expected at 
a hadron collider but
exceeds the sensitivity in the open-top region
at $e^+e^-$ colliders.
\end{abstract}

\newpage
\section{Introduction}
\label{s1}

Among all the fermions included in the Standard Model (SM),
the top quark plays a very unique role.
The mass of top quark is by far the largest and approximates the 
electroweak symmetry breaking scale.
In fact the top quark is the heaviest of all the 
elementary particles discovered up to now.
It means that in the SM Lagrangian the top quark mass
term breaks the $SU(2)_L \times U(1)_Y$ symmetry maximally.
This fact suggests that the top quark couples strongly to the
physics that breaks the electroweak symmetry.
It is therefore important to investigate properties of
top quark in detail, for the purpose of probing the 
symmetry breaking physics as well as
to gain deeper understanding of the origin of the flavor structure.
The standard procedures for investigating top quark properties
are: measurements of fundamental quantities such as its mass
and decay width; 
detailed examinations of various interactions of top quark 
to see if there are signs of new physics.
Among them testing the {\it CP}-violating interactions
of top quark is particularly interesting.
This is because: 
(1) {\it CP}-violation in the top quark sector is extremely small within
the SM.  If any {\it CP}-violating effect is detected in the top sector 
in a near-future experiment, it immediately signals new physics.
(2) There can be many sources of {\it CP}-violation
in models that extend the SM, such as supersymmetric (SUSY) models, 
Leptoquark models (including $R$-parity violating SUSY models), 
multi-Higgs-doublet models,
Extra-dimensions, etc.
(3) In relatively wide class of models
beyond the SM, {\it CP}-violation 
emerges especially sizably in the top quark sector.

Predictions of certain models are as follows.
In the SM, the lowest-order contributions to the 
electric-dipole-moment (EDM) of a quark 
come from three loop diagrams and are 
proportional to $G_F^2 \alpha_s$~\cite{EDM-SM}.  
Assuming that the results for $u$- and $d$-quarks can also be applied to 
the top quark, one may estimate the top quark EDM as
$\sim 10^{-30} e\cm$.
One may also estimate the $Z$-EDM and chromo-EDM of
top quark as $\sim 10^{-30} e\cm$ and 
$\sim 10^{-30} \gs\cm$, respectively, since 
there seems to be no reason that these two EDMs are much suppressed or 
enhanced.
All these EDMs of top quark are quite small 
compared to those corresponding to the ``$\Order{1}$-couplings'',
$e/\mt \sim 10^{-16} e\cm$ and $\gs/\mt \sim 10^{-16} \gs\cm$.  
%This is because the leading contribution is three-loop in SM.  
On the other hand, the top quark EDMs are induced at one loop in many models, 
including multi-Higgs-multiplet models and SUSY models.  
In the two-Higgs-doublet models,
a neutral Higgs $\phi$ can violate {\it CP}
through the Yukawa interaction 
$\psiB (a-\tilde{a}\gamma_5)\psi \, \phi$
\cite{EDM-2HDM,SP92,had-2HDM,epem-2HDM,epem-2HDM-MSSM,epem-gammagamma-2HDM,epem-gammagamma-2HDM-MSSM}.  
The size of the induced EDM is estimated as%
\footnote{
  Here, one power of $\mt$ is necessary to flip chirality.  
  The extra two powers of $\mt$ come from the Yukawa interaction.  
}
$\sim e G_F m_t^3/(4\pi^2 m_\phi^2) = 
3\times10^{-18}e\cm \, (m_\phi/100\GeV)^{-2}$.  
The explicit calculations show somewhat smaller values, 
$10^{-18}\mbox{--}10^{-20}e\cm$, 
depending on $\sqrt{s}$ and $m_\phi$.  
In the Minimal SUSY Standard Model
{\it CP} can be violated in the soft SUSY breaking sector
\cite{epem-2HDM-MSSM,epem-gammagamma-2HDM-MSSM,EDM-MSSM,S92,epem-MSSM}.  
It was shown that the top quark
EDM of $\sim 10^{-19}e\cm$ can be induced by gluino and chargino exchanges, 
assuming a universal gaugino mass and non-universal other soft-breaking 
parameters.  

Present experimental limits on the EDMs of top quark are not stringent
\cite{EDM-present-bound}. 
A limit on the 
$\mbox{chromo-EDM} \lesssim 10^{-16}\gs\cm$
is obtained from the
analyses using $\sigma_{\rm tot}(p\bar{p} \to t\bar{t}X)$ and
$p_{T}$ distributions of prompt photons produced in $qg \to q\gamma$, etc.\ 
at Tevatron and from the analyses using ${\rm Br}(B \to X_s\gamma)$ at CLEO.
The limit on the EDM from the prompt photon distribution is similar: 
$\mbox{EDM} \lesssim 10^{-16}e\cm$.  

There have been a number of sensitivity studies on the
top quark EDMs expected at future hadron colliders and 
in the open-top region ($\sqrt{s} \gg 2\mt$) at
future $e^+e^-$ and $\gamma\gamma$ colliders.  
In hadron collider studies
\cite{SP92,had-2HDM,S92,AAS92,EDM-hadron-col,had-AEell,had-epem},
it is claimed
that with the observables%
\footnote{
  They include the so-called ``optimal observables''~\cite{AAS92,AS92}.  
}
made of elaborated combinations of 
momenta of charged leptons, $b$-quarks, etc., 
experiments at $\sqrt{s} = 500\GeV$ and 
with an integrated luminosity $10\fb^{-1}$ can probe the chromo-EDM down to 
$\mbox{a few}\times 10^{-17}\mbox{--}10^{-18}\gs\cm$, or even to 
$\mbox{a few}\times 10^{-19}\gs\cm$
by raising complexity of the observables.  
However, none of them performs detector simulations, 
which seem to be indispensable for a serious sensitivity study.\footnote{
For instance, it is important to study the effects of mis-assignment
of jets to partons in event reconstructions.
}  
Among several proposed {\it CP}-odd observables, lepton energy asymmetry 
$A_E^\ell = E_{\ell^+}-E_{\ell^-}$ would be the 
simplest one~\cite{SP92,S92,had-AEell}.  
It is claimed that $A_E^\ell$ is sensitive to the imaginary part%
\footnote{
  Note that $A_E^\ell$ probes absorptive part of an amplitude $\calM$, 
  since it is $\CPTT$-odd.  
}
of the chromo-EDM down to $\sim 10^{-18}\gs\cm$ assuming an 
acceptance efficiency $\epsilon = 10\%$.  
Also studies of the EDM and $Z$-EDM of top quark in 
the open-top region at $e^+e^-$ colliders are given in
\cite{epem-2HDM,epem-2HDM-MSSM,epem-gammagamma-2HDM,epem-gammagamma-2HDM-MSSM,epem-MSSM,had-epem,F96,AS92,EDM-epem-col,cr}.
We take as a reference the results of \cite{F96}, which is based on
simulation studies incorporating 
experimental conditions expected at a future
$e^+e^-$ linear collider.
It is shown that, by using the mode 
$t\tB \to b\bar{b}WW \to b\bar{b}q\bar{q}'\ell\nu$ ($\ell = e,\mu$), 
sensitivities to (the real and imaginary parts of) the EDM and $Z$-EDM 
are $\sim 10^{-17}e\cm$ at $\sqrt{s}=500\GeV$, assuming 
an integrated luminosity of $10\fb^{-1}$ and electron beam polarization
of $\pm 80\%$.
%and overall efficiency of the analysis was $\simeq 18\%$.  
% Other studies gave similar number, except those with 
% ``optimal observables''~\cite{AS92}.  
%
Sensitivity to the top chromo-EDM in the open-top region
at $e^+e^-$ colliders is studied 
in~\cite{CEDM-epem-col}; they estimate a sensitivity 
$\sim 10^{-16} \gs\cm$ at
$\sqrt{s}=500\GeV$, assuming an integrated luminosity $50\fb^{-1}$, 
an identification efficiency for top-pair production events $\simeq 100\%$, 
and $E_g^{\rm min}=25\GeV$.  
The last entry is a cut for the minimum gluon-jet energy, on which 
the sensitivity depends crucially.  
No detector simulation is performed in this study.
The sensitivities of $\gamma\gamma$ colliders are studied 
in~\cite{epem-gammagamma-2HDM,epem-gammagamma-2HDM-MSSM,EDM-gamma-gamma-col}; 
they are shown to be similar to those of $e^+e^-$ colliders.  
%-----------------------------------------------------------------------

Certainly
it is desirable to probe the top quark anomalous interactions
at highest possible energy where we have more resolving power,
which motivated the above studies.
On the other hand,
it is known that studying various top quark properties in the
$t\bar{t}$ threshold region at future $e^+e^-$ colliders
is promising and interesting;
particularly the top quark mass will be determined to unmatched
precision.
A number of analyses elucidated physics 
potential of experiments in the $t\bar{t}$ threshold region
\cite{fk,sp,hjk,sfhmn,r11,my,fms,r38,jkp,ps,NNLO,nos,kt}.
Most of them, however, dealt only with the SM interactions.
In this paper we extend these analyses and
study how to probe anomalous {\it CP}-violating
interactions of top quark in the $t\bar{t}$ threshold region.
We note that there are some specific advantages in this region:
\begin{itemize}
\item
The polarization of top quark can be raised to close to
100\% by adjusting longitudinal polarization of $e^-$ beam 
\cite{toppol,r38}.
\item
Since top quarks are produced almost at rest, 
one can reconstruct
the spin information of top quarks from 
distributions of their decay
products without solving detailed kinematics \cite{jkp}.
\item
The QCD interaction is enhanced in this region, so
the cross section is sensitive to the top-gluon ($tg$) couplings.
We can study anomalous $tg$ couplings in a clean environment
in comparison to hadron colliders.
\item
There are less backgrounds from multiple $W$,$Z$ productions
compared to the open-top region.
\item
In certain models (e.g.\ those in which a neutral Higgs is
exchanged between $t$ and $\bar{t}$ \cite{EDM-2HDM}),
the induced top quark EDM and $Z$-EDM are enhanced near
the $t\bar{t}$ threshold.
\end{itemize}
Thus, for the sake of comparison with other kinematical regions,
we would like to know sensitivities to 
{\it CP}-violation achievable
in $e^+e^- \to t\bar{t}$ in the threshold region when these
advantages are taken into account.

In Sec.~\ref{s2} we present a qualitative picture of the
effects of the anomalous {\it CP}-violating interactions 
in the threshold region.
We derive the top quark vertices including the QCD
enhancement in Sec.~\ref{s3}.
The formulas for the polarization vectors and the
spin correlation tensor of $t$ and $\bar{t}$ are 
presented in Sec.~\ref{s4}, followed by their
numerical analyses in Sec.~\ref{s5}.
Sec.~\ref{s6} discusses the observables to be measured
in experiments and gives rough estimates of 
sensitivities to the anomalous couplings.
We summarize and conclude our analyses in Sec.~\ref{s7}.
Some of the notations used in this paper are collected in the
Appendix.

\section{Physical Picture}
\label{s2}

Let us first review the time evolution of 
$t$ and $\bar{t}$, pair-created in $e^+e^-$ collision 
just below threshold, within the SM.
They are created close to each other
at a relative distance $r \sim 1/m_t$ 
and then spread apart non-relativistically.
When their relative distance becomes of the order of
the Bohr radius,
$r \sim ( \alpha_s m_t )^{-1}$,
they start to form a Coulombic boundstate.
When the relative distance becomes 
$r \sim (m_t \Gamma_t )^{-1/2}$,
where $\Gamma_t$ is the decay width of top quark,
either $t$ or $\bar{t}$ decays via electroweak interaction,
and accordingly the boundstate decays.
Numerically these two scales have similar magnitudes,
$( \alpha_s m_t )^{-1} \sim (m_t \Gamma_t )^{-1/2}$,
and are much smaller than the hadronization scale
$\sim 1/\Lambda_{\rm QCD}$.
Since gluons which have wavelengths much longer than the
size of the $t\bar{t}$ system cannot couple to 
this color singlet system, the strong interaction
participating in 
the formation of the boundstate
is dictated by the perturbative domain of QCD.
The spin and $PC$ of the dominantly produced boundstate  
are $J^{PC} = 1^{--}$.
Inside this boundstate: $t$ and $\bar{t}$ are in the 
$S$-wave state ($L = 0$);
the spins of $t$ and $\bar{t}$ are aligned to each other and
pointing to $e^-$ beam direction $\ket{\uparrow\uparrow}$
or to $e^+$ beam direction $\ket{\downarrow\downarrow}$ or
they are in a linear combination of the two states ($S=1$).

\begin{figure}[tbp]
  \hspace*{\fill}
  \begin{picture}(50,60)(0,0)
    \normalsize
    \Line(0,50)(30,50)
    \Line(30,50)(60,50)
    \Line(60,0)(30,0)
    \Line(30,0)(0,0)
    \DashLine(30,50)(30,0){5}
    \BCirc(30,50){4}
    \Text(30,50)[]{$\times$}
    \Text(-4,50)[r]{$t$}
    \Text(-4,0)[r]{$\bar{t}$}
    \Text(34,25)[l]{$g^{\mbox{\tiny C}}$}
  \end{picture}
  \hspace*{\fill}
  \begin{picture}(50,60)(0,0)
    \normalsize
    \Line(0,50)(30,50)
    \Line(30,50)(60,50)
    \Line(60,0)(30,0)
    \Line(30,0)(0,0)
    \DashLine(30,50)(30,0){5}
    \BCirc(30,0){4}
    \Text(30,0)[]{$\times$}
    \Text(-4,50)[r]{$t$}
    \Text(-4,0)[r]{$\bar{t}$}
    \Text(34,25)[l]{$g^{\mbox{\tiny C}}$}
  \end{picture}
  \hspace*{\fill}
  \\
  \hspace*{\fill}
  \caption{\small
The diagrams which contribute to the spin-dependent $CP$-violating
potential between non-relativistic $t$ and $\bar{t}$.
The vertex $\otimes$ represents the {\it CP}-odd interaction of
top quark with gluon; cf.\ eq.~(\ref{effop}).
An exchange of the Coulomb gluon $g^{\mbox{\tiny C}}$ 
gives the leading contribution
of the {\it CP}-odd interaction to the potential.
\label{fig:CEDM}    }
  \hspace*{\fill}
\end{figure}
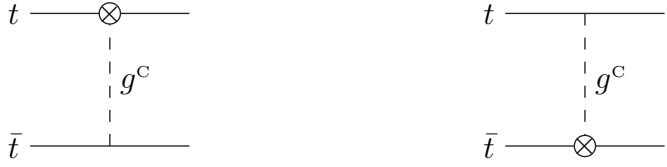

In this paper we consider 
anomalous {\it CP}-violating interactions of top quark with
$\gamma$, $Z$, and $g$.
In particular, we consider
the lowest dimension effective operators which violate {\it CP}:
\bea
{\cal L}_{\mbox{\scriptsize {\it CP}-odd}} =
- \frac{e d_{t\gamma}}{2m_t} 
( \bar{t} i \sigma^{\mu\nu} \gamma_5 t )
\partial_\mu A_\nu
- \frac{g_Z d_{tZ}}{2m_t} 
( \bar{t} i \sigma^{\mu\nu} \gamma_5 t )
\partial_\mu Z_\nu
- \frac{g_s d_{tg}}{2m_t} 
( \bar{t} i \sigma^{\mu\nu} \gamma_5 T^a t )
\partial_\mu G_\nu^a \sepcomma
\nonumber\\
 \sigma^{\mu\nu} \equiv {\textstyle \frac{i}{2}}
[ \gamma^\mu , \gamma^\nu ] \sepcomma
\label{effop}
\eea
where 
$e = g_W \sin \theta_W$ and
$g_Z = {g_W}/{\cos \theta_W}$.
These represent the interactions of $\gamma$, $Z$, $g$ with
the EDM, $Z$-EDM, 
chromo-EDM of top quark, 
respectively.\footnote{
The magnitudes of these EDMs are given by ${e d_{t\gamma}}/{m_t}$,~
${g_Z d_{tZ}}/{m_t}$,~ ${g_s d_{tg}}/{m_t}$,
respectively.
}
Each of these interactions has $C = +1$ and $P = -1$.
We assume that generally the anomalous couplings
$d_{t\gamma}$, $d_{tZ}$, $d_{tg}$
are complex where their imaginary parts may
be induced from some absorptive processes beyond
the SM.
For a non-relativistic $t\bar{t}$ pair produced in
$e^+ e^-$ collision, the 
anomalous couplings of $t\bar{t}$ to $\gamma$
and $Z$ reduce to
\bea
\frac{e d_{t\gamma}}{m_t} 
\mathboldit{A} \cdot \chi^\dagger_{\bar{t}} ( -i \nabla ) \chi_t
+
\frac{g_Z d_{tZ}}{m_t} 
\mathboldit{Z} \cdot \chi^\dagger_{\bar{t}} ( -i \nabla ) \chi_t ,
\label{nr1}
\eea
where $\chi_t$ and $\chi_{\bar{t}}$ denote the 
two-component non-relativistic
fields of $t$ and $\bar{t}$, respectively.
The anomalous top-gluon coupling 
generates effectively a spin-dependent
potential between $t$ and $\bar{t}$
\bea
V_{\mbox{\scriptsize {\it CP}-odd}} = 
\frac{d_{tg}}{m_t} \, 
( \stBI - \stBBI ) \cdot \nabla V_{\rm C}(r) 
\label{cppot}
\eea
through the diagrams shown in Fig.~\ref{fig:CEDM}.
Here, 
$\stBI$ and $\stBBI$ 
denote the spins of non-relativistic
$t$ and $\bar{t}$, respectively; 
$V_{\rm C}(r) = - C_F {\alpha_s}/{r}$
is the Coulomb potential with the 
color factor $C_F = 4/3$.
When $d_{tg}>0$,
the potential $V_{\mbox{\scriptsize {\it CP}-odd}}$ tends to 
align both chromo-EDMs in the direction of chromo-electric field, or,
align $\stBBI$ in the direction of 
$\mathboldit{r} = \mathboldit{r}_t - \bar{\mathboldit{r}}_{t}$
and $\stBI$ in the direction of $- \mathboldit{r}$.

Let us consider effects of these anomalous interactions on the time 
evolution of the $t\bar{t}$ system.
Assuming that the anomalous couplings 
$d_{t\gamma}$, $d_{tZ}$, $d_{tg}$
are small, we consider the effects which arise
in linear perturbation in these couplings.
{\it CP}-violation originating from the $t\gamma$ or $tZ$ coupling
occurs at the stage of the pair creation, 
i.e.\ when $t$ and $\bar{t}$ are
very close to each other.
The generated boundstate has $J^{PC}=1^{+-}$, so
$t$ and $\bar{t}$ are in the $P$-wave ($L=1$) and 
spin-0 state 
$\ket{\uparrow\downarrow}-\ket{\downarrow\uparrow}$.
On the other hand,
{\it CP}-violation 
originating from the $tg$ coupling takes place
after the boundstate formation when 
multiple gluons are exchanged between $t$ and $\bar{t}$,
i.e.\
when $t$ and $\bar{t}$
are separated at a distance of the Bohr radius.
Therefore, first the boundstate is formed in 
$J^{PC}=1^{--}$ ($L=0$ and $S=1$) state and after
interacting via the potential $V_{\mbox{\scriptsize {\it CP}-odd}}$
it turns into $J^{PC}=1^{+-}$
($L=1$ and $S=0$) state.
Since we are interested in the dependences of observables
on the couplings 
$d_{t\gamma}$, $d_{tZ}$, $d_{tg}$ up to linear terms,
we are interested in the interference of the
leading SM amplitude and the amplitude including
these couplings.
The strong phases of these amplitudes
that arise from QCD binding
effects can be calculated reliably using perturbative
QCD.

Which {\it CP}-odd observables are sensitive
to the above {\it CP}-violating couplings?
For the process $e^+e^- \to t\bar{t}$, 
we may conceive of following 
expectation values of combinations of
kinematical variables for
{\it CP}-odd observables:
\bea
&
\left< \,
( \pBI_{e} - \bar{\pBI}_{e} ) \cdot
( \stBI - \stBBI ) 
\, \right> ,
&
\nonumber \\
&
\left< \,
( \ptBI - \ptBBI ) \cdot
( \stBI - \stBBI )
\, \right> ,
&
 \\
&
\left< \,
[ ( \pBI_{e} - \bar{\pBI}_e ) \times
( \ptBI - \ptBBI ) ] \cdot
( \stBI - \stBBI )
\, \right> ,
&
\nonumber
\eea
where the spins and momenta are defined in the
c.m.\ frame.
(The initial state is {\it CP}-even
if we assume the SM interactions of $e^\pm$
with $\gamma$ and $Z$.)
Generally one may think of other combinations involving
$\mathboldit{s}_{e}$ and $\bar{\mathboldit{s}}_{e}$ as well.
However, 
the spin directions of $e^\pm$ are not
independent of their momentum directions for longitudinally
polarized or unpolarized beams.
Therefore, we would like to
measure the difference of the spins
(or the polarization vectors) of $t$ and $\bar{t}$.
Practically we can measure the $t$ and $\bar{t}$ polarization
vectors efficiently using $\ell^\pm$ angular distributions.
It is known that the
angular distribution of the charged lepton $\ell^+$ from the
decay of top quark is
maximally sensitive to the top quark polarization
vector.
In the rest frame of top quark,
the $\ell^+$ angular distribution is given by \cite{langdist}
\bea
\frac{1}{\Gamma_t} \,
\frac{d \Gamma ( t \to b \ell^+ \nu )}{d \cos \theta_{\ell^+}}
= \frac{1 + P \cos \theta_{\ell^+}}{2} 
\label{langdistr}
\eea
at tree level,
where $P$ is the top quark polarization and
$\theta_{\ell^+}$ is the angle of $\ell^+$ measured from the direction
of the top quark polarization vector.\footnote{
Indeed the $\ell^+$ distribution is ideal for extracting
{\it CP}-violation in the $t\bar{t}$ production process;
the above angular distribution is unchanged even if anomalous
interactions are included in the $tbW$ decay vertex,
up to the terms linear in the decay anomalous couplings and within the
approximation $m_b = 0$ \cite{new}.
}
Furthermore, we may think of {\it CP}-odd 
observables bilinear in 
$\stBI$ and $\stBBI$, which
require more complicated analyses for 
their reconstructions from decay products.

We may anticipate following aspects of the {\it CP}-odd
quantity
$\delta \vc{P} = (\vc{P} - \bar{\vc{P}})/2$,
(a half of) the difference of the $t$ and $\bar{t}$
polarization vectors.
It will be directly proportional to the coupling
$d_{t\gamma}$ or $d_{tZ}$ or $d_{tg}$
when only one of the couplings is turned on at a time.
$\delta \vc{P}$ will include a suppression factor
$\beta \simeq |\ptBI|/\mt$, the top quark velocity,
since these couplings 
are accompanied by the top quark momentum; 
cf.\ eqs.~(\ref{nr1}) and (\ref{cppot}).
Thus, this factor will be larger at higher c.m.\ 
energy.
Apart from this $\beta$, 
energy dependence of $\delta \vc{P}$ originating from
the anomalous $tg$ coupling will be different
from that originating from the 
anomalous $t\gamma$ and $tZ$ couplings.
The contribution of the coupling $d_{tg}$ to
$\delta \vc{P}$ will be suppressed 
if the energies of $t$ and $\bar{t}$ are too
large to allow for enhancement of QCD due to
Coulomb binding effects.
This happens if the c.m.\ energy measured from the 
threshold ${\cal E} = \sqrt{s} - 2 m_t$ is much larger
than the Coulomb binding energy,
${\cal E} \gg \alpha_s^2 m_t $.
On the other hand, the contribution of 
$d_{t\gamma}$ or $d_{tZ}$ coupling would not have
such energy dependence since 
{\it CP} violation occurs at the first stage of the
top pair production.
Dependences of $\delta \vc{P}$
on the $e^-$ longitudinal polarization
will be different between the photon-induced effect
and the $Z$-induced effect, since $e_L$ and $e_R$
couple differently to $\gamma$ and $Z$.
These differences in the energy and $e^-$ polarization
dependences can be used to disentangle the effects
of the three anomalous {\it CP} violating interactions
in the $t\bar{t}$ threshold region.\footnote{
Use of the $e^-$ longitudinal polarization for decomposing
the photon-induced effect and the $Z$-induced effect
was advocated in \cite{cr}.
}

\section{The Top Production Vertices}
\label{s3}
\setcounter{footnote}{0}

In this section we present the $t\tB\gamma$ and $t\tB Z$
vertices when the QCD binding effects and the
{\it CP}-violating anomalous couplings are included.
At tree level (and without anomalous interactions), 
the electroweak $t\tB$ vertices are given by
\begin{align}
  \left(\Gamma^{X}\right)^i 
  = v^{tX} \Gamma^i_V - a^{tX} \Gamma^i_A 
  \sepcomma\text{~~~~~~}\quad
  \Gamma^i_V = \gamma^i   \sepcomma\quad
  \Gamma^i_A = \gamma^i\gamma_5  
  \qquad (X = \gamma,Z)
  \sepcomma
\end{align}
times $-ig_X$.  
Since the vertex $(\Gamma^X)^\mu$ is contracted with
the wave functions of $\gamma$ and $Z$ produced by
$e^+e^-$ annihilation, only the space components of
$(\Gamma^X)^\mu$ are relevant.
Here and hereafter, the Latin indices refer to the
space components.
See Appendix for the definition of 
the electroweak couplings
$g_X$, $v^{tX}$ and  $a^{tX}$.  
These vertices are modified by the QCD binding effects and 
by the anomalous interactions as
\begin{align}
  &
  \left(\Gamma^{X}\right)^i 
    = v^{tX} \Gamma^i_V - a^{tX} \Gamma^i_A 
      + d_{tX}\Gamma^i_{X\text{-EDM}}
  \sepcomma
\end{align}
where
\begin{align}
  &
  \Gamma^i_V = 
    \left[
       \left( 1 - \frac{2 C_F \alpha_s}{\pi} \right)
        \gamma^i G(E,p) 
      + i\gamma_5 \frac{p^i}{\mt} \dtg D(E,p)
    \right] \times
    \left( \frac{\pBI^2}{\mt}-(E+i\Gamma_t) \right)
  \sepcomma\nonumber\\
  &
  \Gamma^i_A = 
\left( 1 - \frac{C_F \alpha_s}{\pi}   \right)
    \gamma^i\gamma_5 F(E,p) 
    \times
    \left( \frac{\pBI^2}{\mt}-(E+i\Gamma_t) \right)
  \label{eq:vertex-after-rescat}
  \sepcomma\\
  &
  \Gamma^i_{X\text{-EDM}} = 
    -i\gamma_5 \frac{p^i}{\mt} F(E,p) 
    \times
    \left( \frac{\pBI^2}{\mt}-(E+i\Gamma_t) \right) .
  \nonumber
\end{align}
$\pBI = \ptBI = -\ptBBI$ denotes the 
top quark momentum in the c.m.\ frame,
and $p = |\pBI|$.
We work in the potential-subtracted-mass scheme 
\cite{psmass} instead
of the pole-mass scheme, and 
$E = \sqrt{s} - 2m_{\rm PS}(\mu_f)$ represents
the c.m.\ energy measured from the twice of the 
potential-subtracted mass of top quark;
$m_t$ denotes the pole mass of top quark and it is expressed
in terms of the potential-subtracted mass by
\bea
m_t = m_{\rm PS}(\mu_f) 
+ \frac{C_F \alpha_s(\mu)}{\pi} \, \mu_f
\left[ 1 + \frac{\alpha_s(\mu)}{4\pi} \biggl\{
a_1 - \beta_0 \biggl( \log \frac{\mu_f^2}{\mu^2}-2 \biggr)
\biggr\}
\right] 
\eea
with
\bea
a_1 = \frac{31}{9} C_A - \frac{20}{9} T_F n_f
\sepcomma \quad
\beta_0 = \frac{11}{3}C_A - \frac{4}{3} T_F n_f
\sepcomma
\eea
where $\mu_f$ and $\mu$ denote the renormalization scale of the
potential-subtracted mass and the $\overline{\rm MS}$ coupling,
respectively;
$C_F=4/3$, $C_A=3$, $T_F=1/2$ are the color factors and
$n_f=5$ is the number of active flavors.

$G(E,p)$ and $F(E,p)$ are the $S$-wave and $P$-wave 
Green functions, respectively, defined by 
\bea
  &&\left[- \frac{\nabla^2}{m_t} + V(r;\mu_f) - \left( E+i\Gamma_t \right)
  \right] \tilde{G}(E,\xBI) = \delta^3(\xBI) \sepcomma
\label{sse2}
\\
  &&\left[- \frac{\nabla^2}{m_t} + V(r;\mu_f) - \left( E+i\Gamma_t \right)
  \right] \tilde{F}^k(E, \xBI) = -i \partial^k \delta^3(\xBI)
  \sepcomma
  \label{pwse2}
\eea
and
\bea
{G}(E,p) &=& \int d^3\xBI \, e^{\scriptsize -i\pBI \cdot \xBI } \, 
\tilde{G}(E,\xBI) \sepcomma
\\
p^k \,
F(E,p) &=& \int d^3\xBI \, e^{\scriptsize -i\pBI \cdot \xBI } \, 
\tilde{F}^k(E,\xBI) \sepperiod
\label{ftpgf2}
\eea
$V(r;\mu_f)$ is the Fourier transform of the two-loop
renormalization-group-improved QCD potential, where
the infrared renormalon-pole is subtracted and absorbed
into the definition of $m_{\rm PS}(\mu_f)$.
See \cite{nos} for details.
One may also write conveniently as
\begin{align}
&&  G(E,p) = \langle \pBI |
\frac{1}{{\pBI^2}/{\mt}+V-(E+i\Gamma_t)}
| \xBI=\vec{0} \rangle  
  \sepcomma\quad
\label{braketG}
\\
&&  \pBI F(E,p) = \langle \pBI |
\frac{1}{{\pBI^2}/{\mt}+V-(E+i\Gamma_t)} \pBI
| \xBI=\vec{0} \rangle  
  \sepperiod
\label{pF}
\end{align}
The Green function associated with the gluon anomalous
coupling is given by 
\bea
D(E,p) = G(E,p) - F(E,p) .
\label{greend}
\eea

We can see from eqs.~(\ref{eq:vertex-after-rescat}) that 
the effects of all the anomalous {\it CP}-violating interactions 
are suppressed by $|\pBI|/\mt \simeq \beta$.  
Thus, for consistency we have incorporated all
$\Order{\alpha_s}=\Order{\beta}$ corrections
in the SM vertices.

%For $V \to 0$ the Green functions reduce to
%
%\begin{align*}
%  G(E,p) ,~~ F(E,p) \to 
%  1 / \left( \frac{\pBI^2}{\mt} - (E+i\Gamma_t) \right) ,
%~~~~~
%D(E,p) \to 0
%.
%\end{align*}
%
%Thus in the weak potential limit, or at $\sqrt{s}\gg 2\mt$ where 
%the QCD binding effects diminish, 
%the vertices reduce to the tree-level ones for the
%$t\gamma$ and $tZ$ anomalous 
%interactions whereas the effect of $tg$ anomalous
%coupling vanishes.

A sketch of derivations of these vertices goes as follows.
Using non-relativistic forms of the 
$t$ and $\bar{t}$ propagators, 
\begin{align}
  &  S_F(k^\mu+\frac{q^\mu}{2})
  \simeq \frac{1+\gamma^0}{2} \frac{i}{{\cal E}/2 + k^0 - |\kBI|^2/(2\mt)+i\Gamma_t/2}
  \sepcomma\nonumber\\
  &  S_F(k^\mu-\frac{q^\mu}{2})
  \simeq \frac{1-\gamma^0}{2} \frac{i}{{\cal E}/2 - k^0 - |\kBI|^2/(2\mt)+i\Gamma_t/2}
  \sepcomma
\end{align}
where $q^\mu = (2m_t+{\cal E},\vec{0})$ and $k^\mu = (k^0,\kBI)$
in the c.m.\ frame,
a self-consistent equation for the vector vertex,
similar to that given in \cite{sp}, reads
\begin{align}
  \Gamma_V^i({\cal E},p^\mu )
  &=
  \gamma^i
  + C_F (-i\gs)^2 \int\!\!\!\frac{\diffn{k}{4}}{(2\pi)^4}
    \frac{i}{{\cal E}/2 + k^0 - |\kBI|^2/(2\mt)+i\Gamma_t/2}
    \times
  \nonumber\\
  &\qquad{}
    \times
    \frac{i}{{\cal E}/2 - k^0 - |\kBI|^2/(2\mt)+i\Gamma_t/2}
    \times
  \nonumber\\
  &\qquad{}
    \times
    \biggl[ \quad
      \gamma^0\frac{1+\gamma^0}{2} \Gamma_V^i({\cal E},k^\mu ) 
      \frac{1-\gamma^0}{2}\gamma^0
  \nonumber\\
  &\qquad\qquad{}
    - \pfrac{\dtg}{2\mt} \sigma^{j0}\gamma_5 {(p-k)^j}
      \frac{1+\gamma^0}{2} \Gamma_V^i({\cal E},k) \frac{1-\gamma^0}{2}\gamma^0
  \nonumber\\
  &\qquad\qquad{}
    + \pfrac{\dtg}{2\mt} \gamma^0\frac{1+\gamma^0}{2} \Gamma_V^i({\cal E},k) 
      \frac{1-\gamma^0}{2} \sigma^{j0}\gamma_5 {(p-k)^j}
  \nonumber\\
  &\qquad{}
    \biggr]
    \times \frac{i}{|\pBI-\kBI|^2}
  \sepperiod
  \label{sceq}
\end{align}
Since there is no $p^0$-dependence on the right-hand-side, 
consistency requires 
$\Gamma_V^i({\cal E},p^\mu ) = \Gamma_V^i({\cal E},\pBI)$.  
Thus we can trivially integrate over $p^0$ and obtain 
\begin{align}
  \Gamma_V^i({\cal E},\pBI)
  &=
  \gamma^i
  - \int\!\!\!\frac{\diffn{\kBI}{3}}{(2\pi)^3} 
    \frac{-\CF\gs^2}{|\pBI-\kBI|^2}
    \frac{1}{|\kBI|^2/\mt-({\cal E}+i\Gamma_t)}
    \times
\nonumber  \\
  &\qquad{}
    \times
    \frac{1+\gamma^0}{2} \left[
      \Gamma_V^i({\cal E},\kBI)
    - \frac{\dtg}{2\mt} {(p-k)^j}
      \left\{ 
          \sigma^{j0}\gamma_5 \Gamma_V^i({\cal E},\kBI)
        + \Gamma_V^i({\cal E},\kBI) \sigma^{j0}\gamma_5
      \right\}
    \right] \frac{1-\gamma^0}{2} 
  \sepperiod\nonumber
\\
\end{align}
We decompose the vertex function 
$\Gamma_V^i({\cal E},\pBI)$ into different spinor
structures as
\begin{align}
%  &\Shift{-3em}
  \frac{1+\gamma^0}{2} \Gamma_V^i({\cal E},\pBI) \frac{1-\gamma^0}{2}
%  \nonumber\\
  &=
  \frac{1+\gamma^0}{2} \biggl[ \quad
    \gamma^i \Gamma_G({\cal E},p)
+ \gamma^j 
\Bigl( \frac{p^ip^j}{|\pBI|^2} - \frac{1}{3}\delta^{ij} \Bigr)
\Gamma_B({\cal E},p)
  \nonumber\\
  &\qquad\qquad{}
    + \gamma^i \gamma_5 \Gamma_F({\cal E},p)
    + i\gamma_5 \frac{p^i}{\mt} \Gamma_D({\cal E},p)
    \biggr] \frac{1-\gamma^0}{2} .
\end{align}
By pluging this expression into the integral equation above, 
one obtains integral equations for scalar functions 
$\Gamma_G({\cal E},p)$, etc.  
One can see that $\Gamma_D({\cal E},p) = \Order{\dtg}$, 
$\Gamma_B({\cal E},p) = \Order{\dtg^{~2}}$.  
Thus we neglect $\Gamma_B({\cal E},p)$ hereafter.
Let us write  
\begin{align}
  &  \Gamma_G({\cal E},p) = 
    \left( \frac{\pBI^2}{\mt} - ({\cal E}+i\Gamma_t) \right) 
    G({\cal E},p)
  \sepcomma\nonumber\\
  &  \Gamma_F({\cal E},p) = 
    \left( \frac{\pBI^2}{\mt} - ({\cal E}+i\Gamma_t) \right) 
    F({\cal E},p)
  \sepcomma\\
  &  \Gamma_D({\cal E},p) = 
    \left( \frac{\pBI^2}{\mt} - ({\cal E}+i\Gamma_t) \right) 
    \dtg D({\cal E},p)
  \sepperiod\nonumber
\end{align}
Then $G$, $F$ and $D$ satisfy
\begin{align}
  &  \left( \frac{\pBI^2}{\mt} - ({\cal E}+i\Gamma_t) \right) G({\cal E},p)
  + \int\!\!\!\frac{\diffn{\kBI}{3}}{(2\pi)^3}
    \VT_\rmC(|\pBI-\kBI|) G({\cal E},k) 
  = 1
  \sepcomma
  \\
  &  \left( \frac{\pBI^2}{\mt} - ({\cal E}+i\Gamma_t) \right) p^i F({\cal E},p)
  + \int\!\!\!\frac{\diffn{\kBI}{3}}{(2\pi)^3} 
    \VT_\rmC(|\pBI-\kBI|) k^i F({\cal E},k) 
  = p^i
  \sepcomma
\\
  &  \left( \frac{\pBI^2}{\mt} - ({\cal E}+i\Gamma_t) \right) p^i D({\cal E},p)
  + \int\!\!\!\frac{\diffn{\kBI}{3}}{(2\pi)^3} 
    \VT_\rmC(|\pBI-\kBI|) k^i D({\cal E},k) 
\nonumber \\
& ~~~~~~~~~~~~~~~~~~~~~~~ ~~~~~~~~~~ ~~~
  = \int\!\!\!\frac{\diffn{\kBI}{3}}{(2\pi)^3} 
    \VT_\rmC(|\pBI-\kBI|) (k-p)^i G({\cal E},k) 
\end{align}
with $\VT_\rmC(q) = - C_F 4\pi\alpha_s/q^2$.
Comparing the third equation with the first two equations,
we find that $D = G - F$.  
The first two equations are equivalent to 
eqs.~(\ref{sse2})-(\ref{pF}) apart from the fact that
the Coulomb potential $-C_F \alpha_s/r$
is replaced by\footnote{
The replacement is justified: In Coulomb gauge the
${\cal O}(\alpha_s)$ corrections to the potential come
solely from the vacuum polarization of Coulomb-gluon
\cite{feinberg}.
Hence, the net effect is to replace the
fixed-coupling constant $\alpha_s(\mu)$ in the leading-order
by the V-scheme
running coupling constant $\alpha_V(|\pBI-\kBI|)$.
} 
$V(r;\mu_f)$ and that
the renormalization-group-improved
potential-subtracted-mass scheme \cite{nos}
is used instead of the pole-mass scheme.

The axial-vector vertex was derived 
in \cite{r11}:
\begin{align}
  \Gamma_A^i   \simeq 
  \left( \frac{p^2}{\mt} - (E+i\Gamma_t) \right) 
  \gamma^i\gamma_5 F(E,p)
  \sepperiod
\end{align}
The hard-vertex factor for the vector vertex was
derived in \cite{kk} and that for the axial-vector vertex
in \cite{kz}.
We may also derive $\Gamma^i_{X\text{-EDM}}$ in a
similar manner.

Two comments would be in order here.
One might think that including the non-renormalizable
interactions eq.~(\ref{effop}) into loop integrals 
[e.g.\ eq.~(\ref{sceq})] causes
ultra-violet divergences and leads to unpredictability.
We note that only non-relativistic domains of the
loop integrals are relevant
in resummations of the Coulomb singularities.
In fact high momentum regions are
effectively cut off in the self-consistent equations
due to our non-relativistic approximation.
Thus, we can calculate unambiguously the leading contributions
of these effective interactions.
In this regard, in eqs.~(\ref{eq:vertex-after-rescat})
the hard-vertex correction
factors are associated only with the SM
contributions.  
We cannot determine hard-vertex corrections to the
vertices including the anomalous interactions since
non-renormalizability of these interactions 
matters at this order.

The simple form of the Green function including the $tg$ anomalous
interaction eq.~(\ref{greend}) is a consequence of the
following fact.
The effect of
$V_{\mbox{\scriptsize {\it CP}-odd}}$
integrated over the time interval from $t=0$ to $t=T$ can be written as
$({d_{tg}}/{m_t}) \, ( \stBI - \stBBI ) \cdot 
[ \pBI(0) - \pBI(T) ]$ using the equation of motion 
$\dot{\pBI} = - \nabla V_{\rm C}$.
Namely the difference of the top quark momenta at $t=0$ and
at $t=T$ carries the net effect of the chromo-electric field which
aligns the EDMs during $0<t<T$.
Concisely, for $H = \pBI^2/m_t + V_{\rm C}$ and 
$\delta H = V_{\mbox{\scriptsize {\it CP}-odd}}$,
the variation of the time evolution of the $t\bar{t}$ system is
expressed as
\bea
\delta \left( e^{-i H T} \right) =
i \left[ \, \frac{d_{tg}}{m_t} \, ( \stBI - \stBBI ) \cdot \pBI 
\sepcomma \,
e^{-i H T} \, \right] \sepperiod
\eea
Thus, the propagation at a fixed energy is given by
$(i{d_{tg}}/{m_t}) \, ( \stBI - \stBBI ) \cdot \pBI ( G - F )$;
cf.\ eqs.~(\ref{braketG}) and (\ref{pF}).

\section{The Polarization Vectors and the Spin
         Correlation Tensor of $t$ and $\bar{t}$}
\label{s4}

Using the vertices derived in the previous section
we may write down 
the production cross section of a $t\bar{t}$ pair
in the threshold region.
The cross section, where ($t$,$\bar{t}$) have
momenta ($\ptBI$,$-\ptBI$) and the spins $+\frac{1}{2}$
along the quantization axes ($\stBI$,$\stBBI$)
in the c.m.\ frame, is given by
\begin{align}
  \frac{\diffn{\sigma}{}(\stBI,\stBBI)}{\diffn{\pBI_t}{3}}
  =   \frac{\diffn{\sigma}{}}{\diffn{\pBI_t}{3}}
  \frac{%
      1
    + \dprod{\PolBR}{\stBI}
    + \dprod{\PolBBR}{\stBBI}
    + (\stBI)_i(\stBBI)_j\PolQBR_{ij} }%
  {4}
  \sepperiod
  \label{spinprojectedcs}
\end{align}
Here, $|\stBI|=|\stBBI|=1$.
On the right-hand-side,
$d\sigma/\diffn{\pBI_t}{3}$ represents 
the production cross section when the spins of
$t$ and $\bar{t}$ are summed over:
\begin{align}
  \frac{\diffn{\sigma}{}}{\diffn{\pBI_t}{3}}
  &=
\left( 1 - \frac{4C_F\alpha_s}{\pi} \right)
  \frac{N_C \alpha^2 \Gamma_t}{2\pi m_t^4}
  \frac{1-\Pol_{e^+}\Pol_{e^-}}{2}
\nonumber \\
  & \times
  |G(E,p_t)|^2 (a_1+\chi a_2) \left\{
    1 + 2{\rm Re}\left[ \CFB\frac{F(E,p_t)}{G(E,p_t)} \right]
\beta \cos\theta_{te}
  \right\}
  \sepcomma
\label{unpolcs}
\end{align}
where $\beta = |\ptBI|/m_t$ and 
$\cos\theta_{te} = \dprod{\peBI}{\ptBI}/(|\peBI|\,|\ptBI|)$; 
$\alpha$ is the fine structure constant;
$N_c = 3$ is the number of colors;
$\chi$ is a 
function of the initial $e^\pm$ longitudinal
polarizations $P_{e^\pm}$:
\bea
\chi = \frac{ P_{e^+} - P_{e^-} }{ 1 - P_{e^+} P_{e^-} } 
\sepperiod
\label{chi}
\eea 
If the positron beam is unpolarized ($P_{e^+} = 0$), 
$\chi = - P_{e^-}$.
The coefficient $\CFB$ and the constants $a_1$, $a_2$
are defined below.  

In eq.~(\ref{spinprojectedcs})
$\PolBR$ and $\PolBBR$ represent the polarization vectors 
of $t$ and $\tB$, respectively.  
Both the SM and the anomalous interactions contribute to 
the polarizations: 
\begin{align}
  \PolBR  = \PolBR_\rmSM  + \delta\PolBR  \sepcomma\quad
  \PolBBR = \PolBBR_\rmSM + \delta\PolBBR  
  \sepperiod
  \label{ttbarpol}
\end{align}
The SM contributions are {\it CP}-even (except for tiny
{\it CP}-violating effects which we neglect) and are
equal for $t$ and $\bar{t}$.
On the other hand, the anomalous {\it CP}-odd contributions
are opposite in sign:
\begin{align}
  \PolBBR_\rmSM =   \PolBR_\rmSM  \sepcomma\quad
  \delta\PolBBR = - \delta\PolBR  \sepperiod
\end{align}
Note that we are working up to linear terms in
the anomalous couplings.  
Hereafter we express these vectors by components: 
\begin{align}
  \PolBR = \Polpara\nBIpara + \Polperp\nBIperp + \Polnorm\nBInorm
  \sepcomma\quad
  \stBI = \spara\nBIpara + \sperp\nBIperp + \snorm\nBInorm
  \sepcomma
\end{align}
where the orthonormal basis is defined from the $e^-$ beam 
direction and the top momentum direction:
\begin{align}
  \nBIpara \equiv \frac{\peBI}{|\peBI|}  
  \sepcomma\quad
  \nBInorm \equiv \frac{\cprod{\peBI}{\ptBI}}{|\cprod{\peBI}{\ptBI}|}  
  \sepcomma\quad
  \nBIperp \equiv \cprod{\nBInorm}{\nBIpara}
  \sepperiod
\end{align}
Then the polarization of $t$ is given by
\begin{align}
  & \PolSMpara
  = \Cpara^0 + \pRe{\Cpara^{1} \frac{F}{G}} \beta \cos\theta_{te}
  \sepcomma\\
  & \PolSMperp
  = \pRe{\Cperp \frac{F}{G}} \beta \sin\theta_{te}
  \sepcomma\\
  & \PolSMnorm
  = \pIm{\Cnorm \frac{F}{G}} \beta \sin\theta_{te}
\end{align}
for the contributions from the SM interactions%
\footnote{
  These results were derived in~\cite{r38,jkp,ps}.  
}, and 
\begin{align}
  & \delta\Polpara = 0
\label{delpolpara}
  \sepcomma\\
  & 
  \delta\Polperp
  =
  \left[
      \pIm{ \Bperp^g d_{tg} \frac{D}{G}}
    + \pIm{\Bperp^{\gamma} d_{t\gamma} \frac{F}{G}}
    + \pIm{\Bperp^{Z}      d_{tZ}      \frac{F}{G}}
  \right] \beta \sin\theta_{te}
\label{delpolperp}
  \sepcomma\\
  &
  \delta\Polnorm
  =
  \left[
      \pRe{\Bnorm^g d_{tg} \frac{D}{G}}
    + \pRe{\Bnorm^{\gamma} d_{t\gamma} \frac{F}{G}}
    + \pRe{\Bnorm^{Z}      d_{tZ}      \frac{F}{G}}
  \right] \beta \sin\theta_{te}
\label{delpolnorm}
\end{align}
for the contributions from the anomalous interactions.  
The coefficients $\Cpara^0$ etc.\ are defined below.  

There is also a term bilinear in $\stBI$ and $\stBBI$, which
represents the correlation of $t$ and $\bar{t}$ spins:
\begin{align}
  (\stBI)_i(\stBBI)_j\PolQBR_{ij}
  =
    (\stBI)_i(\stBBI)_j 
\left( {\PolQBR_{ij}}_{\rmSM}
  + \delta\PolQBR_{ij} \right)
  \sepcomma
\end{align}
where
\begin{align}
  (\stBI)_i(\stBBI)_j {\PolQBR_{ij}}_{\rmSM}
  &=
    \spara\sBpara
  + (\spara\sBperp+\sperp\sBpara) \pRe{\Cnorm \frac{F}{G}}
    \beta \sin\theta_{te}
  \nonumber \\
  &\qquad{}
  + (\spara\sBnorm+\snorm\sBpara) \pIm{\Cperp \frac{F}{G}}
    \beta \sin\theta_{te}
  \sepcomma
\end{align}
and
\begin{align}
  (\stBI)_i(\stBBI)_j \, & \delta\PolQBR_{ij}
  =
  \nonumber \\
  &
  (\spara\sBperp-\sperp\sBpara)\left[
        \Im\left(\Bnorm^g d_{tg} \frac{D}{G}\right)
      + \Im\left(\Bnorm^{\gamma} d_{t\gamma} \frac{F}{G}\right)
      + \Im\left(\Bnorm^{Z}      d_{tZ}      \frac{F}{G}\right)
    \right]
    \beta \sin\theta_{te}
  \nonumber \\
  + & (\spara\sBnorm-\snorm\sBpara)\left[
      \Re\left(\Bperp^g d_{tg} \frac{D}{G}\right)
    + \Re\left(\Bperp^{\gamma} d_{t\gamma} \frac{F}{G}\right)
    + \Re\left(\Bperp^{Z}      d_{tZ}      \frac{F}{G}\right)
    \right]
    \beta \sin\theta_{te}
\label{delQ}
  \sepperiod
\end{align}

The coefficients $C_i$'s and  $B^X_i$'s
included in $\PolBR_\rmSM$, $\delta\PolBR$,
${\PolQBR_{ij}}_{\rmSM}$, $\delta\PolQBR_{ij}$
are defined as follows.
For the SM contributions,%
\footnote{
  Our notations are similar to those of \cite{jkp,ps}.  
  There are two differences: (i) our $a_3$ and $a_4$ 
  are factor two smaller than theirs;
  (ii) our $C_N$ is defined in opposite sign to theirs.
}
\begin{align}
  &  \Cpara^0(\chi) = - \frac{a_2+\chi a_1}{a_1+\chi a_2} 
  \sepcomma\quad
  && \Cpara^1(\chi) = 2\Cpara^0\Cnorm+\Cperp
       = 2(1-\chi^2)\frac{a_2 a_3-a_1 a_4}{(a_1+\chi a_2)^2} 
  \sepcomma\nonumber\\
  &  \Cperp(\chi) = - \frac{a_4+\chi a_3}{a_1+\chi a_2} 
  \sepcomma\quad
  && \Cnorm(\chi) = \frac{a_3+\chi a_4}{a_1+\chi a_2}
       = \CFB 
\sepperiod
  \label{eq:Cnorm}
\end{align}
The coefficients for the {\it CP}-odd contributions are given by
\begin{align}
  & \Bperp^\gamma(\chi) \, {d_{t\gamma}} 
         + \Bperp^Z(\chi) \, {d_{tZ}} 
 = \frac{a_5+\chi a_6}{a_1+\chi a_2} 
\sepcomma \nonumber \\
  &  \Bnorm^\gamma(\chi) \, {d_{t\gamma}} + \Bnorm^Z(\chi) \, {d_{tZ}}
= \frac{a_6+\chi a_5}{a_1+\chi a_2}
         \label{eq:Bnorm}
  \sepcomma
\end{align}
and
\begin{align}
  \Bperp^g(\chi) &= -1 
  \sepcomma\nonumber\\[2ex]
  \Bperp^\gamma(\chi)
  &= \frac{1}{a_1+\chi a_2}
    \left\{
        \left( [v^e v^t]^*\,{v^{e\gamma}} + [a^e v^t]^*\,{a^{e\gamma}} \right)
      + \chi
        \left( [v^e v^t]^*\,{a^{e\gamma}} + [a^e v^t]^*\,{v^{e\gamma}} \right)
    \right\}
  \nonumber\\
  &= \frac{1}{a_1+\chi a_2}
    \left(
        [v^e v^t]^*\,{v^{e\gamma}} + \chi [a^e v^t]^*\,{v^{e\gamma}}
    \right)
  \sepcomma\nonumber\\[2ex]
  \Bperp^Z(\chi)
  &= \frac{1}{a_1+\chi a_2}
    \left\{
        \left( [v^e v^t]^*\,{v^{eZ}} + [a^e v^t]^*\,{a^{eZ}} \right)
      + \chi
        \left( [v^e v^t]^*\,{a^{eZ}} + [a^e v^t]^*\,{v^{eZ}} \right)
    \right\}
    d(s)
  \sepcomma\nonumber\\[2ex]
  \Bnorm^g(\chi)
  &= \Cpara^0(\chi)
  \label{eq:BnormZ}
  \sepcomma\\[2ex]
  \Bnorm^\gamma(\chi)
  &= \frac{1}{a_1+\chi a_2}
    \left\{
        \chi
        \left( [v^e v^t]^*\,{v^{e\gamma}} + [a^e v^t]^*\,{a^{e\gamma}} \right)
      + \left( [v^e v^t]^*\,{a^{e\gamma}} + [a^e v^t]^*\,{v^{e\gamma}} \right)
    \right\}
  \nonumber \\
  &= \frac{1}{a_1+\chi a_2}
    \left(
        \chi [v^e v^t]^*\,{v^{e\gamma}} + [a^e v^t]^*\,{v^{e\gamma}}
    \right)
  \sepcomma\nonumber\\[2ex]
  \Bnorm^Z(\chi)
  &= \frac{1}{a_1+\chi a_2}
    \left\{
        \chi
        \left( [v^e v^t]^*\,{v^{eZ}} + [a^e v^t]^*\,{a^{eZ}} \right)
      + \left( [v^e v^t]^*\,{a^{eZ}} + [a^e v^t]^*\,{v^{eZ}} \right)
    \right\}
    d(s) \sepperiod
  \nonumber
\end{align}
Symbols $a_{1\sim 6}$ denote combinations of the 
electroweak parameters:
\begin{align}
     a_1 &= \left|[v^e v^t]\right|^2+\left|[a^e v^t]\right|^2 
  \sepcomma\quad
  &  a_2 &= 2 \pRe{[v^e v^t]^*\,[a^e v^t]} 
  \label{eq:def-a1}
  \sepcomma\\
     a_3 &= [v^e v^t]^*\,[a^e a^t] + [a^e v^t]^*\,[v^e a^t] 
  \sepcomma\quad
  &  a_4 &= [v^e v^t]^*\,[v^e a^t] + [a^e v^t]^*\,[a^e a^t] 
  \sepcomma\nonumber
\end{align}
and
\begin{align}
  a_5 
  &= [v^e v^t]^*\,[v^e d^t] + [a^e v^t]^*\,[a^e d^t] 
  \nonumber\\
%  &= [v^e v^t]^*\,{v^{e\gamma}} + [a^e v^t]^*\,{a^{e\gamma}} 
%      {d_{t\gamma}} 
%    + [v^e v^t]^*\,{v^{eZ}} + [a^e v^t]^*\,{a^{eZ}} 
%      d(s)\,{d_{tZ}} 
%  \nonumber\\
  &= [v^e v^t]^*\,{v^{e\gamma}} {d_{t\gamma}} 
    + [v^e v^t]^*\,{v^{eZ}}d(s)\,{d_{tZ}} + [a^e v^t]^*\,{a^{eZ}} 
      d(s)\,{d_{tZ}} 
  \sepcomma\nonumber\\[2ex]
  a_6 
  &= [v^e v^t]^*\,[a^e d^t] + [a^e v^t]^*\,[v^e d^t] 
  \label{eq:def-a5}
  \\
%  &=  [v^e v^t]^*\,{a^{e\gamma}} + [a^e v^t]^*\,{v^{e\gamma}} 
%      {d_{t\gamma}} 
%    + [v^e v^t]^*\,{a^{eZ}} + [a^e v^t]^*\,{v^{eZ}} 
%      d(s)\,{d_{tZ}}
%  \nonumber\\
  &=  [v^e v^t]^*\,{a^{eZ}}d(s)\,{d_{tZ}}
    + [a^e v^t]^*\,{v^{e\gamma}} {d_{t\gamma}}  + [a^e v^t]^*\,{v^{eZ}} 
      d(s)\,{d_{tZ}} \sepperiod
  \nonumber
\end{align}
Symbols $[v^e a^t]$ etc.\ are defined in the Appendix,
eq.~(\ref{eq:eff-coupling}), 
and $d(s)$ is the ratio of the $Z$ propagator to 
the $\gamma$ propagator [eq.~(\ref{eq:ds-def=prop-ratio})].  
%These combinations $a_{1\sim 6}$ are chosen so that the 
%dependence on the initial polarizations can be seen easily;
%each pair of $(a_1,a_2)$, $(a_3,a_4)$, $(a_5,a_6)$ is ``conjugate'' 
%to each other with respect to exchange of the $e^+e^-$ couplings.
%

We comment here why there is no $\nBIpara$ component in 
$\delta \vc{P}$ [eq.~(\ref{delpolpara})]
or why there are only a few 
components in $\delta\PolQBR_{ij}$ [eq.~(\ref{delQ})].
$\delta \vc{P}$ and $\delta\PolQBR_{ij}$ originate
from interferences of the leading SM amplitude $M_0$
and the amplitude proportional to the anomalous
couplings $\delta M$.
The SM amplitude $M_0$ is in a linear combination
of spin $S_{\!\Parallel}=\pm 1$ states 
($c_+ \ket{\uparrow\uparrow} 
+ c_- \ket{\downarrow\downarrow}$),
whereas the {\it CP}-reversed amplitude $\delta M$ is in
spin $S_{\!\Parallel}=0$ state 
($\ket{\uparrow\downarrow} - \ket{\downarrow\downarrow}$);
see Sec.~\ref{s2}.
In order to produce a non-zero interference between the
two amplitudes, either one of 
the spins of $t$ and $\bar{t}$ must be
flipped.
This is possible only by sandwitching the spin operator
$\hat{S}_{\!\perp}^{(i)}$ or $\hat{S}_{\!\mathrm{N}}^{(i)}$
($i = t$ or $\bar{t}$).
%Thus, the absence of $\nBIpara$ components is due to the
%selection rules with respect to eigenvalues of 
%$\hat{S}_{\!\Parallel}$.

We can understand from symmetry considerations
the combinations of the electroweak couplings and the
Green functions in each term of the 
production cross section
${\diffn{\sigma}{}(\stBI,\stBBI)}/{\diffn{\pBI_t}{3}}$.
This provides a non-trivial cross check of the formulas
presented in this section;
the argument is presented in \cite{nagano-dthesis}.

\section{Numerical Analyses of $\delta \vc{P}$
         and $\delta\PolQBR_{ij}$}
\label{s5}

In this section we study numerically the
polarization vectors and the spin correlation tensor
derived in the previous section.
We use the input parameters:
$m_{\rm PS}(\mu_f)=175$~GeV,
$\mu_f = 3$~GeV, $\mu = 20$~GeV,
$m_Z = 91.19$~GeV,
$\alpha_s(m_Z) = 0.118$,
and $\sin^2 \theta_W = 0.2312$.

\begin{figure}[tbp]
  \hspace*{\fill}
  \begin{minipage}{5.0cm}\centering
    \hspace*{-1cm}
    \includegraphics{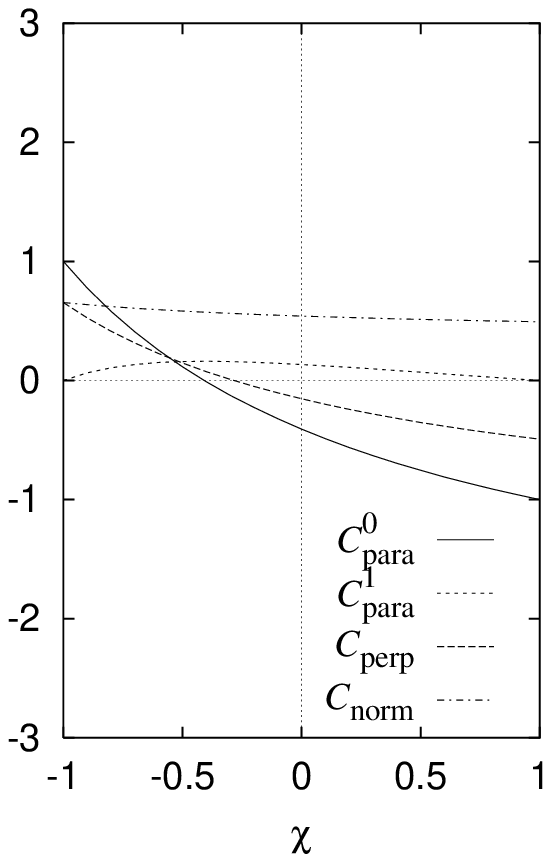}
    (a)
  \end{minipage}
  \hspace*{\fill}
  \begin{minipage}{5.0cm}\centering
    \hspace*{-1cm}
    \includegraphics{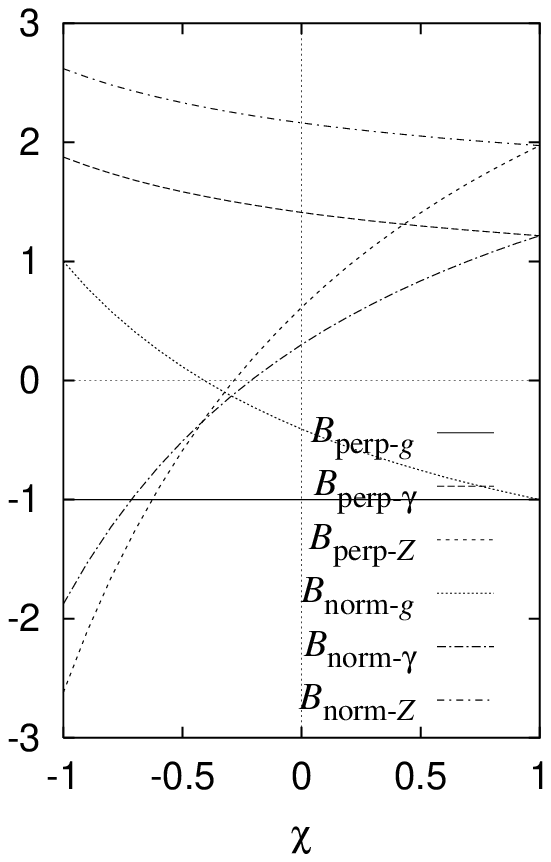}
    (b)
  \end{minipage}
  \hspace*{\fill}
  \\
  \hspace*{\fill}
  \begin{Caption}\caption{\small
      The electroweak coefficients $C_i$'s and $B_i^X$'s
(for $\PolBR$, $\PolBBR$ and $\PolQBR_{ij}$)
vs.\ the initial $e^\pm$ polarization parameter
      $\chi$.  
In the figures, 
      $C_{\rm para}=\Cpara$, $C_{\rm perp}=\Cperp$, 
$C_{\rm norm}=\Cnorm$, etc. 
      (a) The coefficients for the SM contributions.  
      (b) The coefficients for the contributions from the
          anomalous couplings.  
      \label{fig:Cpara}
  }\end{Caption}
  \hspace*{\fill}
\end{figure}

First we examine the coefficients 
$C_i$'s and  $B^X_i$'s, which represent
combinations of electroweak couplings.
They are given as a function of $\chi$; 
cf.~eq.~(\ref{chi}).
Fig.~\ref{fig:Cpara}(a) shows the coefficients for the
SM contributions $\PolBR_\rmSM$ and
${\PolQBR_{ij}}_{\rmSM}$.
Except for $\Cpara^1$, typical sizes of the coefficients 
are order one.
Fig.~\ref{fig:Cpara}(b) shows the coefficients for the
{\it CP}-violating contributions $\delta\PolBR$ and
$\delta\PolQBR_{ij}$.
Typical sizes of all these coefficients 
are order one.
We see that their dependences on $\chi$ are
different.

\begin{figure}[tbp]
  \hspace*{\fill}
  \begin{minipage}{8cm}\centering
    \includegraphics[width=8cm]{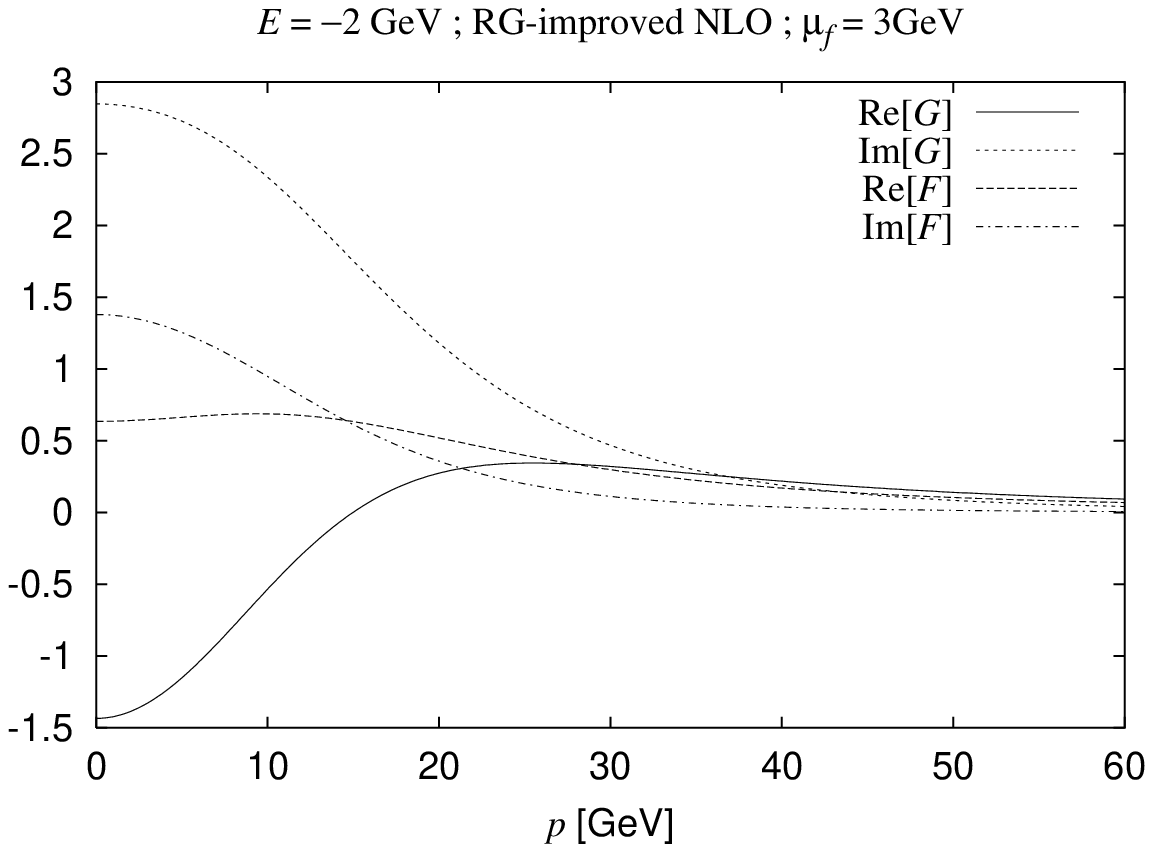}
    (a)
  \end{minipage}
  \hspace*{\fill}
  \begin{minipage}{8cm}\centering
    \includegraphics[width=8cm]{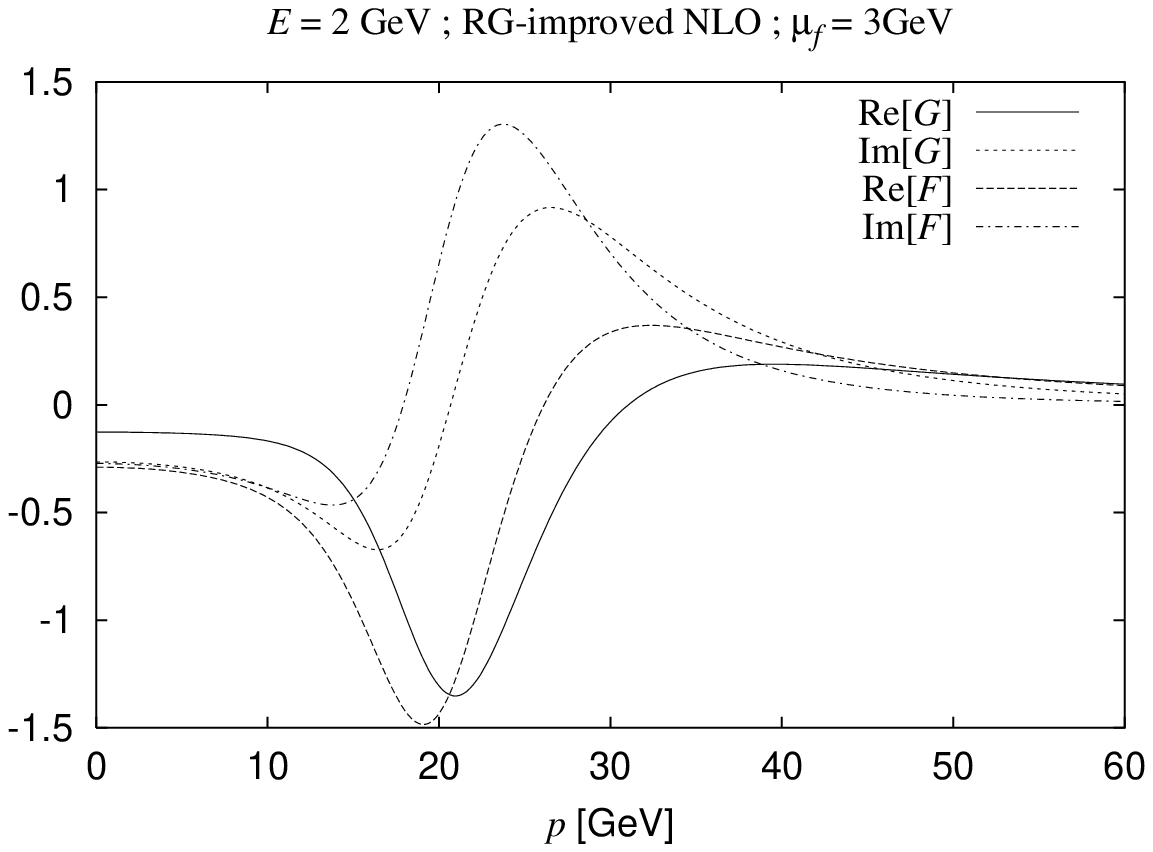}
    (b)
  \end{minipage}
  \hspace*{\fill}
  \\
  \hspace*{\fill}
  \begin{Caption}\caption{\small
The $S$-wave and $P$-wave Green functions vs.\
the top quark momentum $p$ at
(a) $E = -2$~GeV, (b) $E = +2$~GeV.
      \label{FG}
  }\end{Caption}
  \hspace*{\fill}
\end{figure}

\begin{figure}[tbp]
  \hspace*{\fill}
  \begin{minipage}{8cm}
    \includegraphics[width=8cm]{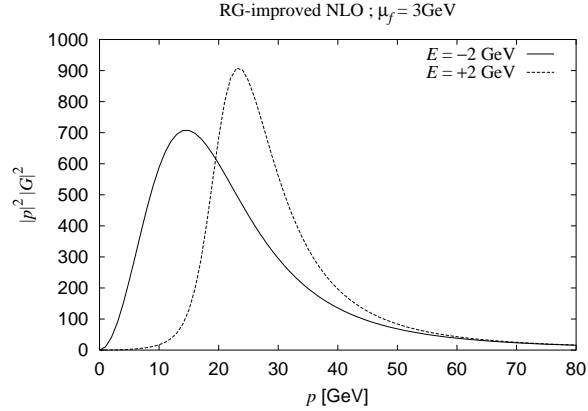}
  \end{minipage}
  \hspace*{\fill}
  \\
  \hspace*{\fill}
  \begin{Caption}\caption{\small
$p^2 |G(E,p)|^2$ vs.\ the top quark momentum $p$ at fixed
c.m.\ energies.
These are proportional to the leading-order momentum distributions 
of top quark.
      \label{momdist}
  }\end{Caption}
  \hspace*{\fill}
\end{figure}

\begin{figure}[tbp]
  \hspace*{\fill}
  \begin{minipage}{8cm}
    \includegraphics[width=8cm]{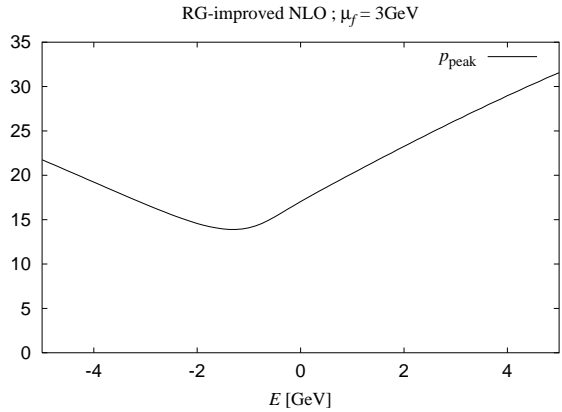}
  \end{minipage}
  \hspace*{\fill}
  \\
  \hspace*{\fill}
  \begin{Caption}\caption{\small
      The peak momentum $p_{\rm peak}$ of the momentum distribution 
      $\diffn{\sigma}{}/\diffn{p}{} \propto |\pBI|^2 |G|^2$
      vs.\ the c.m.\ energy measured from the twice of the
      potential-subtracted mass, $E = \sqrt{s}-2m_{\rm PS}(\mu_f)$.  
      It represents the typical momentum of top quark as a
      function of $E$.
      \label{fig:FT_scanE_p_peak}
  }\end{Caption}
  \hspace*{\fill}
\end{figure}

\begin{figure}[tbp]
  \hspace*{\fill}
  \begin{minipage}{8cm}\centering
    \includegraphics[width=8cm]{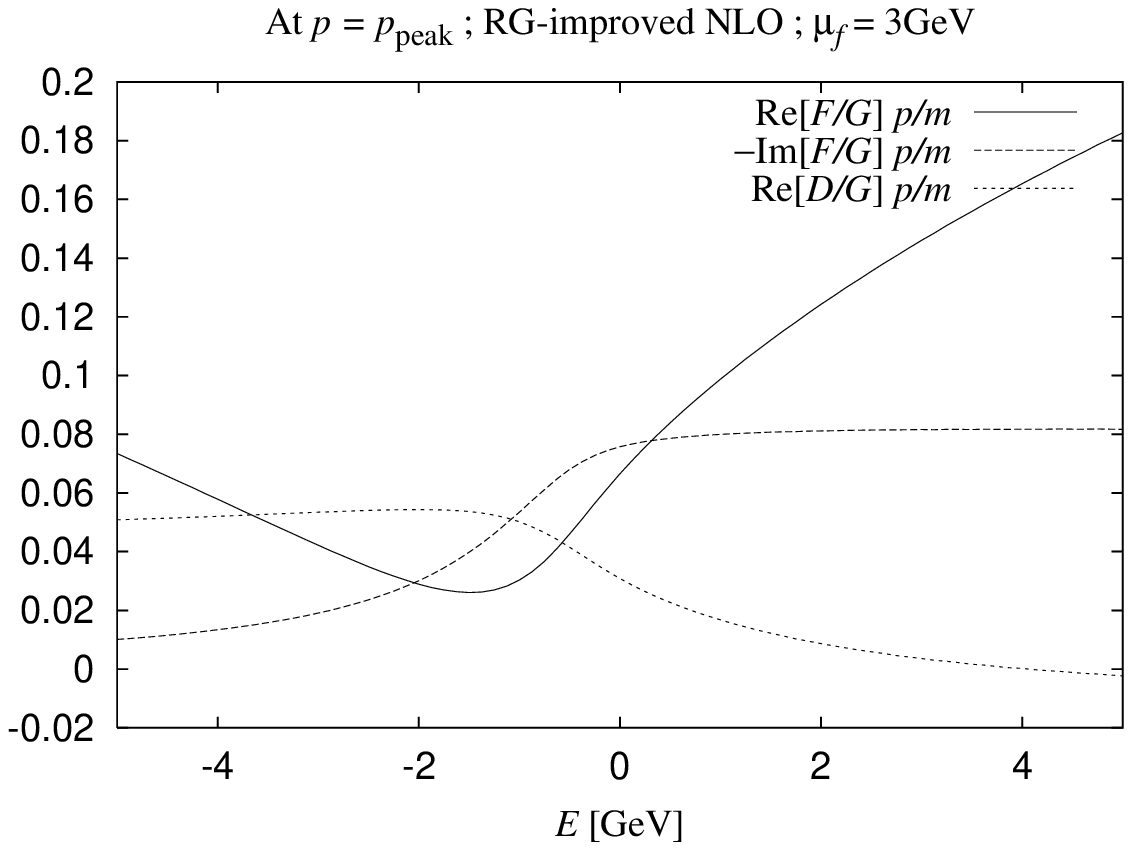}
    (a)
  \end{minipage}
  \hspace*{\fill}
  \begin{minipage}{8cm}\centering
    \includegraphics[width=8cm]{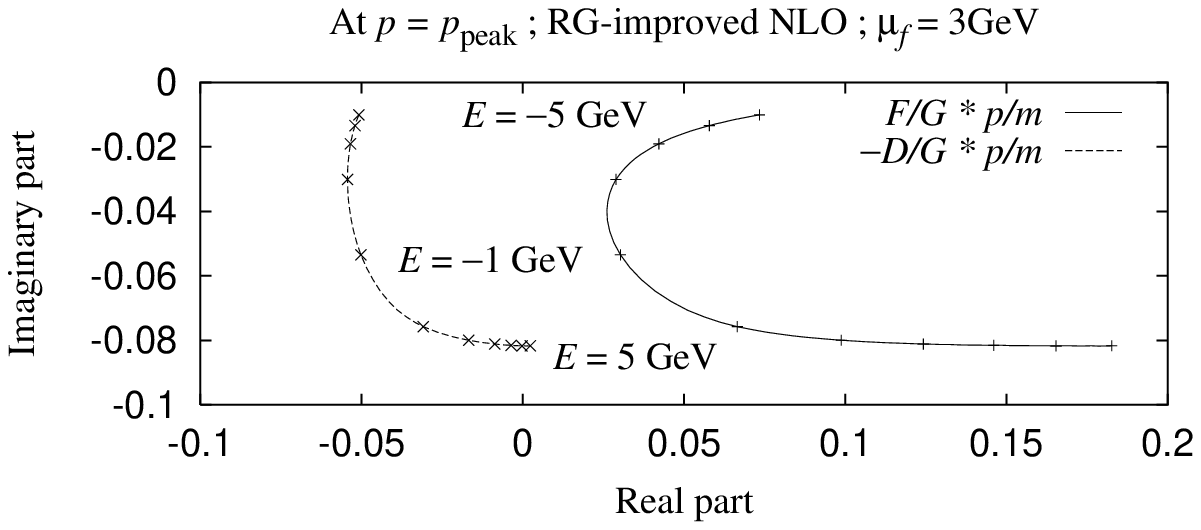}
    (b)
  \end{minipage}
  \hspace*{\fill}
  \\
  \hspace*{\fill}
  \begin{Caption}\caption{\small
      The ratios of the Green functions times the ``velocity'' of 
      top quark evaluated at the 
      peak momentum $p_{\rm peak}$ of the momentum distribution 
      $\diffn{\sigma}{}/\diffn{p}{}$; see Fig.~\ref{fig:FT_scanE_p_peak}.
      (a) These are given as a function of $E$.
      (b) These are plotted on a
      complex plane as $E$ is varied.
      \label{fig:FT_scanE_b}
  }\end{Caption}
  \hspace*{\fill}
\end{figure}

Next we examine the $S$-wave and $P$-wave Green functions.
In Figs.~\ref{FG}(a) and (b) are shown the Green functions at
$E = -2$~GeV and $E = +2$~GeV, respectively.
They depend on both the energy $E$ and the top momentum
$p_t$.
The momentum distribution of top quark
$d\sigma/\diffn{\pBI_t}{3} \propto p_t^2 | G(E,p_t) |^2$
has a peak ($p_t = p_{\rm peak}$) at a given c.m.\ energy \cite{sfhmn},
see Figs.~\ref{momdist} and \ref{fig:FT_scanE_p_peak}.
Then we may plot the ratios $\beta F/G$ and $\beta D/G$,
included in $\delta \vc{P}$ and $\delta\PolQBR_{ij}$,
as a function of the energy $E$ alone by choosing
the top momentum to be the peak momentum $p_{\rm peak}$.
These are shown in Figs.~\ref{fig:FT_scanE_b}.
We see that the size of $|\beta F/G|$ is 5--20\% while 
the size of $|\beta D/G|$ is 5--10\%.
Clearly their energy dependences are different.
Also it can be seen that the strong phases are quite sizable.

One may understand these behaviors of the Green functions
semi-quantitatively using analytic formulas.
The relation
\begin{align}
  p_{\rm peak} \simeq \left| \sqrt{m_t\,(E+1\GeV+i\Gamma_t)} \right|
  \label{eq:p_peak_appr}
\end{align}
agrees qualitatively with Fig.~\ref{fig:FT_scanE_p_peak},
in particular at $E>0$.\footnote{
For $V \to 0$, 
$p_{\rm peak} = \left| \sqrt{m_t\,(E+i\Gamma_t)} \right|$
holds exactly.
}  
Here, $1\GeV \simeq 2m_t-M_{1S} = \mbox{``binding energy''}$.  
For a stable quark pair with the Coulomb potential, $G$ and $F$ 
can be obtained analytically for on-shell kinematics~\cite{fkk}: 
\begin{align}
  &  \lim_{\stackrel{\scriptstyle \Gamma_t \to 0}
                    {E \to p_t^2/m_t}              }
  \left( E - \frac{\ptBI^2}{\mt} + i\Gamma_t \right) G(E,p_t) 
\biggr|_{V=V_{\rm C}}
  = \exp\!\pfrac{\pi p_\rmB}{2p_t}\,
    \Gamma\!\left(1+i\frac{p_\rmB}{p_t}\right)
  \sepcomma\\
  &  \lim_{\stackrel{\scriptstyle \Gamma_t \to 0}
                    {E \to p_t^2/m_t}              }
  \left( E - \frac{\ptBI^2}{\mt} + i\Gamma_t \right) F(E,p_t) 
\biggr|_{V=V_{\rm C}}
  = \left( 1-i\frac{p_\rmB}{p_t} \right) \, \exp\!\pfrac{\pi p_\rmB}{2p_t}\,
    \Gamma\!\left(1+i\frac{p_\rmB}{p_t}\right)
  \sepcomma
\end{align}
where $p_\rmB \equiv C_F \alpha_s m_t/2 \simeq 20\GeV$.  
Thus, we may find a sensible approximation formula
\begin{align}
  \left. \frac{F}{G} \right|_{p=p_{\rm peak}}
  \simeq 1-i\frac{p_\rmB}{\sqrt{m_t\,(E+1\GeV+i\Gamma_t)}}
  \sepperiod
\end{align}
This agrees qualitatively well with Figs.~\ref{fig:FT_scanE_b}.  
It follows that $D/G \to 0$ and $F/G \to 1$ when 
$|E + i\Gamma_t| \gg \alpha_s^2 m_t$.
%Note that there is no energy dependence of $\bRe{F/G}$ if 
%$\Gamma_t = 0$.  

Combining the analyses of the electroweak coefficients and
the Green functions, we find that the typical sizes of
the {\it CP}-odd quantities $\delta \vc{P}$ and 
$\delta\PolQBR_{ij}$ are 5--20\% times the couplings
$( \dtp , \dtZ , \dtg )$ in the threshold region.
Using the different dependences on
the $e^\pm$ polarizations and on the c.m.\ energy,
we will be able to disentangle the effects of the
three anomalous couplings in the
$t\bar{t}$ threshold region.  
A more comprehensive numerical study of the coefficients
and the Green functions is presented in \cite{nagano-dthesis}.

\section{Observables and Sensitivity Estimates}
\label{s6}

So far we have considered the {\it CP}-odd quantities
related to the spins of $t$ and $\bar{t}$.
These quantities are, however, not directly measurable
observables in experiments.
In this section we focus on $\delta \vc{P}$
and consider how to extract it.
Then we give rough estimates of
sensitivities to the {\it CP}-violating couplings
$d_{t\gamma}$, $d_{tZ}$, $d_{tg}$ expected at future $e^+ e^-$
colliders.

The top quark polarization vector can be extracted most efficiently
using the angular distribution of charged leptons from the decay of
top quarks.
The charged lepton angular distribution
in the $t\bar{t}$ threshold region, in the leading order,
is given by 
\bea
\frac{{\rm d}\sigma(e^+e^- \! \to t\bar{t} \to b\ell^+\nu\bar{b}W^-)}
{{\rm d}^3\ptBI {\rm d}\Omega_\ell}
\simeq
\frac{{\rm d}\sigma}
{{\rm d}^3\ptBI} \times
%( 1 + \delta_0 + \delta_1 \cos \theta_{te} )
%\nonumber \\
%\times
\frac{1}{\Gamma_t} \,
\frac{{\rm d}\Gamma_{t \to b\ell^+\nu}
(\vc{P}) }
{{\rm d}\Omega_\ell} \sepperiod
%\times \left[ 1 +
%\xi (|\ptBI|,E,E_l,\cos\theta_{lt}) \right] .
%\nonumber \\
\label{factorize}
\eea
The left-hand-side shows that this is 
the differential cross section where the three-momentum of 
parent top quark and the direction of charged lepton are fixed,
while all other variables are integrated over.
The right-hand-side shows that it is given as
a product of the $t\bar{t}$ production cross section
eq.~(\ref{unpolcs}) and the decay angular distribution
from free polarized top quarks.
The top polarization vector is given by eq.~(\ref{ttbarpol}).
The lepton angular distribution in the top rest frame is given
by eq.~(\ref{langdistr}).
It coincides with the angular distribution in the laboratory 
frame in the leading order,
since top quarks are almost at rest in the threshold region.
Hence, the expectation
value of the lepton three-momentum projected onto an
arbitrary chosen direction $\vc{n}$ 
is proportional to the top quark polarization vector 
in the same approximation \cite{jkp}:
\bea
\langle \langle \vc{n} \cdot \pBI_{\ell} \rangle \rangle 
\simeq
\frac{1 + 2r + 3r^2}{12(1+2r)} \, m_t \times
\vc{n} \cdot
\vc{P}
\sepcomma
\label{expec}
\eea
where $r = m_W^2/m_t^2$, and
$\langle \langle \cdots \rangle \rangle$ denotes
an average taken for a fixed top three-momentum $\ptBI$.

Taking a {\it CP}-odd combination, the contributions of
the anomalous interactions can be extracted as
\bea
\langle \langle \vc{n} \cdot ( \pBI_{\ell} + \pBBI_{\ell} )  \rangle \rangle 
\simeq
\frac{1 + 2r + 3r^2}{6(1+2r)} \, m_t \times
\vc{n} \cdot \delta \vc{P}
\sepperiod
\label{cpoddleptonmom}
\eea
By choosing $\vc{n} = \nBIperp$ and $\nBInorm$, we
can extract the components eqs.~(\ref{delpolperp}) and
(\ref{delpolnorm}) of $\delta \vc{P}$.
The above formula remains valid
even if we include the full ${\cal O}(\alpha_s)$
corrections (in particular the final-state interactions)
in the SM parts of eqs.~(\ref{factorize}) \cite{ps}
and (\ref{expec}) \cite{jkp}, since the
pure SM contributions drop in the {\it CP}-odd 
combination.
Alternatively, we may consider a slightly different
observable
\bea
\langle \langle \vc{n} \cdot 
( \vc{n}_{\ell} + \bar{\vc{n}}_{\ell} )  \rangle \rangle 
\simeq
\frac{2}{3} \, 
\vc{n} \cdot \delta \vc{P}
\sepcomma
\label{expec-angle}
\eea
where $\vc{n}_{\ell}$/$\bar{\vc{n}}_{\ell}$ denote the directions
of $\ell^\pm$.
This observable would be useful if the $\ell^\pm$ directions
can be measured more accurately than their three-momenta,
such as in the case of $\tau^\pm$.

In experiments the lepton-plus-4-jet mode can be used to reconstruct
the lepton momentum and the top quark three-momentum 
simultaneously \cite{fms}.
In order to detect a signal of {\it CP}-violation, 
it is not necessary to reconstruct
the top quark three-momentum with a high accuracy.
One should define the top quark momentum merely as the sum of 
all the visible momenta in a top-jet cluster, so no stringent cuts
are required to reduce missing momentum.
The only important point is that any experimental cut should be imposed in a
{\it CP} symmetric way.
Later when we measure accurately the values of the couplings
$d_{t\gamma}$, $d_{tZ}$, $d_{tg}$, we would need to 
reconstruct the top quark three-momentum to a reasonable accuracy.

In order to extract $\delta\PolQBR_{ij}$ we need to measure
spin correlations of $t$ and $\bar{t}$.
Instead of eq.~(\ref{factorize}) we should consider a double
differential decay distribution of $t$ and $\bar{t}$,
which can be obtained using the formula of \cite{tsai}.
We may think of observables such as
$\langle ( \pBI_t - \pBBI_t ) \cdot ( \pBI_\ell \times \pBBI_\ell )
\rangle$ for {\it CP}-odd observables sensitive to $\delta\PolQBR_{ij}$.
Here, we do not discuss extraction of $\delta\PolQBR_{ij}$
any further and leave the subject to our future work.

Let us make rough estimates of sensitivities to the
{\it CP}-violating couplings $d_{t\gamma}$, $d_{tZ}$, $d_{tg}$
expected in future experiments.
Eq.~(\ref{expec-angle}) shows that a statistical reconstruction of the
top quark polarization using 
lepton directions is quite efficient.
The top quark
polarization vector projected to a certain direction 
${\Pol}=\mean{\stBI \cdot \vc{n}}$ is
given by 
\begin{align}
  {\Pol} \simeq
  \frac{N_{\uparrow}-N_{\downarrow}}{N_{\uparrow}+N_{\downarrow}}
  \sepcomma
\end{align}
where $N_{\uparrow}$ ($N_{\downarrow}$) denotes the number of
top quarks with spin in the direction $\vc{n}$ ($-\vc{n}$).
Hence, the statistical error of ${\Pol}$ may be estimated as 
$\delta^{\rm (stat)} {\Pol} \sim 1/\sqrt{N_{\rmeff}}$, where
$N_{\rmeff}$ stands for the number of events used for the analysis. 
Assuming an integrated luminosity $\intlum = 50\fb^{-1}$ 
and a detection efficiency $\epsilon = 0.6$, 
\begin{align}
  N_{\rmeff} 
  &= 
  \sigma_{t\bar{t}} \times 
  \intlum \times (2 B_{\ell} B_{h}) \times \epsilon
  \nonumber\\
  & \simeq
  0.5\, {\rm pb} 
\times 50\fb^{-1} \times \left(2\cdot\frac{2}{9} \cdot \frac{2}{3}\right)
  \times 0.6
%  \nonumber\\
  = 4\times 10^3 \text{\,events,}
\end{align}
which means $1/\sqrt{N_{\rm eff}} \simeq 1.5 \times 10^{-2}$.  
The leptonic (hadronic) branching fraction $B_{\ell}$ ($B_{h}$) of $W^\pm$ 
into $e^\pm$ and $\mu^\pm$ (hadrons) is given by $2/9$ ($2/3$).  

Using the relations
\begin{align}
  \delta \Polperp
  = 
  \pIm{\Bperp^g d_{tg}\frac{D}{G}} \beta \sin\theta_{te}
  \sepcomma \quad
  \delta \Polnorm
  = 
  \pRe{\Bnorm^g d_{tg}\frac{D}{G}} \beta \sin\theta_{te}
  \sepcomma
\end{align}
the statistical error of $\dtg$ is estimated to be
\begin{align}
  \delta^{\rm (stat)} d_{tg}
  &\sim
  \frac{1}{|{B^g_i {D}/{G}}| \beta}
  \times \frac{\dint{-1}{1}\diffn{(\cos\theta_{te})}{}\,1}
    {\dint{-1}{1}\diffn{(\cos\theta_{te})}{}\,\sin\theta_{te}}
  \times \frac{1}{\sqrt{N_{\rmeff}}}
  \nonumber\\
  &\simeq
  \frac{1}{0.1} \times \frac{4}{\pi} \times 1.5 \times 10^{-2}
  \nonumber\\
  &\simeq
  0.2 \sim \Order{10\%} \sepperiod
\end{align}
Note that we have similar sensitivities to both the real part and
imaginary part of $d_{tg}$ independently using the two components
of the top quark polarization vector.
The above value translates to a sensitivity to the
chromo-EDM of top quark at
\begin{align}
\delta^{\rm (stat)} \left(  
\frac{g_s}{m_t} d_{tg} \right)
\sim 10^{-17} g_s\cm
  \sepperiod
\end{align}
Since all the electroweak coefficients ($\Cperp$, $\Bnorm$, etc.) 
are of similar sizes, 
and so are the ratios of the Green functions ($\pIm{F/G}$, etc.), 
sensitivities to the EDM and $Z$-EDM are estimated to be
at the same order: 
\begin{align}
\delta^{\rm (stat)} \left(  
  \frac{e}{m_t} d_{t\gamma} \right)
\sim 10^{-17} e\cm  \sepcomma\quad
\delta^{\rm (stat)} \left(  
  \frac{\gZ}{m_t} \dtZ \right) 
\sim 10^{-17} \gZ\cm  \sepperiod
\end{align}

A Monte Carlo simulation study 
is also in progress incorporating realistic
experimental conditions expected at a future
$e^+e^-$ collider \cite{topWG}.  
They show that high detection efficiency is possible with 
simple event selection criteria and $b$-tagging.  
Up to now, only the lepton energy asymmetry 
$\mean{E_{\ell^+}-E_{\ell^-}}$ 
for the dilepton-plus-2-jet events was studied \cite{ACFA-proc}.
The 1$\sigma$ statistical error corresponding to $100\fb^{-1}$ 
was obtained as
\begin{align}
  \delta^{\rm(stat)} \left[ \mean{E_{\ell^+}-E_{\ell^-}} \right]
  = 0.65\GeV  \sepperiod
\end{align}
They studied the bounds on the anomalous couplings setting
the input values to be $\dtp = \dtZ = \dtg = 0$.
Based on our calculations, they obtained 
$|{\rm Re}[e^{i\phi_\gamma}{\dtp}]| < 1.5$, 
$|{\rm Re}[e^{i\phi_Z}{\dtZ}]| < 1.0$, 
$|{\rm Re}[e^{i\phi_g}{\dtg}]| < 3.9$ at 95\% confidence-level 
(statistical errors only),
where $e^{i\phi_X}$'s are the relevant strong phases.
%
%\begin{align}
%  |{\dtp}| < 1.48  \sepcomma\quad
%  |{\dtZ}| < 1.04  \sepcomma\quad
%  |{\dtg}| < 3.93   \quad
%\mbox{ (95\% CL) }
%\end{align}
%
%where only the statistical errors were estimated.
%These bounds correspond to 
%
%\begin{align}
%  &  \Bigl|\frac{e}{m_t} d_{t\gamma}\Bigr| 
%     < 4.8 \times 10^{-18} e\cm   \sepcomma\\
%  &  \Bigl|\frac{\gZ}{m_t} \dtZ\Bigr|
%     < 3.4 \times 10^{-18} \gZ\cm \sepcomma\\
%  &  \Bigl|\frac{g_s}{m_t} d_{tg}\Bigr|
%     < 1.2 \times 10^{-17} \gs\cm \sepperiod
%\end{align}
%
In fact, the 
lepton energy asymmetry is not a good observable for extracting 
$\delta \vc{P}$ in the threshold region.
It is suppressed by
$\beta\sim 10\%$ compared to the {\it CP}-odd combination of
the lepton three-memontum, eq.~(\ref{cpoddleptonmom}).
Moreover the branching fraction for the lepton-plus-4-jet mode is 
larger than that for the dilepton-plus-2-jet mode.  
Thus, we expect that
the sensitivities to the anomalous EDMs will be better
by a factor 10 or more
if we use the lepton three-momentum or the lepton direction.  
This is consistent with the naive estimates we made above.  

\section{Summary and Conclusions}
\label{s7}

In this paper we studied how to probe the anomalous 
{\it CP}-violating couplings of top quark with $\gamma$,
$Z$ and $g$ in the $t\bar{t}$ threshold region at
future $e^+e^-$ colliders.
The anomalous couplings contribute to the difference
of the $t$ and $\bar{t}$ polarization vectors,
$\delta \vc{P} = (\vc{P}-\bar{\vc{P}})/2$,
as well as to the spin correlation tensor $\delta\PolQBR_{ij}$.
We studied dependences of these {\it CP}-odd quantities on the
$e^\pm$ beam polarizations, c.m.\ energy, and top quark momentum.
We find that the typical sizes of $\delta \vc{P}$ and 
$\delta\PolQBR_{ij}$ are 5--20\% times the couplings
$( \dtp , \dtZ , \dtg )$ in the threshold region.
Experimentally we can measure $\delta \vc{P}$ efficiently using the
expectation value of the {\it CP}-odd combination of the
$\ell^\pm$ momenta, $\pBI_\ell + \pBBI_\ell$, or of the $\ell^\pm$ directions,
$\vc{n}_{\ell} + \bar{\vc{n}}_{\ell}$.
We have similar sensitivities to both the real part and
imaginary part of $\dtp$, $\dtZ$, $\dtg$
independently using the two components
of the top quark polarization vector $\delta \Polperp$ and
$\delta \Polnorm$.
Taking advantage of 
different dependences of $\delta \vc{P}$ on
the $e^\pm$ polarizations and on the c.m.\ energy,
we will be able to disentangle the effects of the
three anomalous couplings $\dtp$, $\dtZ$, $\dtg$ in the
$t\bar{t}$ threshold region.  
We made rough estimates of sensitivities to the anomalous
couplings expected at future $e^+e^-$ colliders,
considering as a simplest example extraction
of $\delta \vc{P}$ from the $\ell^\pm$ distributions.
For an integrated luminosity of $50~{\rm fb}^{-1}$,
we estimated
\bea
\delta^{\rm (stat)} d_{t\gamma}, \, \,
\delta^{\rm (stat)} d_{tZ}, \, \,
\delta^{\rm (stat)} d_{tg}
\sim \Order{10\%} 
\eea
when only one of the couplings is turned on at a time.\footnote{
We would be able to improve the sensitivities by using other
observables:
$\delta \vc{P}$ can be extracted also from distributions of charm
quarks from $W^\pm$ instead of $\ell^\pm$;
$\delta \vc{Q}_{ij}$ can be extracted using correlations of
$\ell^\pm$, $b$, $c$ distributions.
}
The above values translate to sensitivities to the top quark
EDM, $Z$-EDM and chromo-EDM:
\bea
\begin{array}{c}
\delta^{\rm (stat)} \left(  
{\displaystyle
  \frac{e}{m_t} d_{t\gamma} 
}\right)
\sim 10^{-17} e\cm  \sepcomma\quad
\delta^{\rm (stat)} \left(  
{\displaystyle
  \frac{\gZ}{m_t} \dtZ 
}\right) 
\sim 10^{-17} \gZ\cm  
\sepcomma\\
\delta^{\rm (stat)} \left(  
{\displaystyle
\frac{g_s}{m_t} d_{tg} 
}\right)
\sim 10^{-17} g_s\cm \sepperiod
\end{array}
\eea
The sensitivities to the top quark EDM and $Z$-EDM are comparable
to those attainable in the open-top region at $e^+e^-$
colliders \cite{F96}.
The sensitivity to $\dtg$ is worse than that expected at 
a hadron collider 
\cite{SP92,had-2HDM,S92,AAS92,EDM-hadron-col,had-AEell,had-epem} 
but exceeds the sensitivity in the open-top region
at $e^+e^-$ colliders \cite{CEDM-epem-col}.
We note that there is an advantage of the $t\bar{t}$ 
threshold region.
The clean environment of an $e^+e^-$ collider
enables accurate determination of the value of the top-gluon
anomalous coupling $d_{tg}$
if its value happens to be larger than $\Order{10\%}$.
On the other hand, 
at hadron colliders it would be difficult to measure
the value of the coupling with a similar accuracy even if
a {\it CP}-violating effect is detected.
Regarding energy upgrading scenario of a future linear
$e^+e^-$ collider, it is possible that the machine
operates first in the $t\bar{t}$ threshold region for a significant
amount of time,
while measuring the top quark mass precisely, etc., before
the beam energies will be increased to the open-top region.
Therefore it would be desirable that measurements of the anomalous
couplings can be performed concurrently with 
other unique measurements near threshold,
with sensitivities comparable to those in the open-top region.
Unfortunately the sensitivities to the {\it CP}-violating
couplings achievable in the $t\bar{t}$
threshold region are one to three orders of magnitude
larger than the predicted sizes of top quark EDMs in the
models reviewed in Sec.~\ref{s1}.
Using the results of this work,
a Monte Carlo study incorporating realistic experimental conditions
expected at a future $e^+e^-$ linear collider is underway \cite{topWG}.

\section*{Acknowledgements}

We are grateful to the hospitality and stimulating atmosphere
at the Summer Institute '99 (August 1999, Yamanashi, 
Japan) where this work
was initiated.
We thank K.~Fujii, Z.~Hioki, K.~Ikematsu and J.H.~K\"uhn
for valuable discussions.
One of the authors (Y.S.) would like to thank 
S.~Rindani, T.~Takahashi, M.~Tanabashi and M.~Yamaguchi 
for useful discussions.
This work was supported in part
by the Japan-German Cooperative Science
Promotion Program.

\begin{appendix}

\section*{Appendix: Conventions and Notations}
\label{app}

In $e^+ e^- \to t\tB$, 
both $\gamma$ and $Z$ are exchanged in $s$-channel.  
Their effects can be combined in terms of effective couplings.  
We denote the SM vertices for electron and top quark by
\begin{align}
  {\Lambda_X}_\mu = 
     g_{\!\mbox{\tiny $X$}} 
     \left(v^{eX}\gamma_\mu - a^{eX}\gamma_\mu\gamma_5 \right)
  \sepcomma\qquad
  \Gamma_X^\mu = 
     g_{\!\mbox{\tiny $X$}} 
     \left( v^{tX}\gamma^\mu - a^{tX}\gamma^\mu\gamma_5 \right)
  \qquad
  (X = \gamma,Z) \sepcomma
\end{align}
where 
\begin{gather}
  g_\gamma = e = g \sin\thetaW \sepcomma\quad
  v^{f\gamma} = Q_f  \sepcomma\quad
  a^{f\gamma} = 0  \sepcomma\nonumber\\
  \gZ = \frac{g}{\cos\thetaW} \sepcomma\quad
  v^{fZ} = \frac{1}{2} T_{3L} - Q_f \sin^2\theta_W  \sepcomma\quad
  a^{fZ} = \frac{1}{2} T_{3L}  
  \label{eq:def=pZ-coupl}
  \sepperiod
\end{gather}
The amplitude for $e^+ e^- \to t\tB$ at tree level 
can be written as
\begin{align}
%  &
%  \frac{1}{s}
%  \bigl( \vB(\pB_e) {\Lambda_\gamma}_\mu u(p_e) \bigr)
%  \bigl( \uB(p_t) \Gamma_\gamma^\mu v(\pB_t) \bigr)
%  +
%  \frac{1}{s-m_Z^2}
%  \bigl( \vB(\pB_e) {\Lambda_Z}_\mu u(p_e) \bigr)
%  \bigl( \uB(p_t) \Gamma_Z^\mu v(\pB_t) \bigr)
%  \\
%  =
  &
  \sum_{X = \gamma,Z}
  \frac{1}{s-m_X^2}
  \bigl( \vB(\pB_e) {\Lambda_X}_\mu u(p_e) \bigr)
  \bigl( \uB(p_t) \Gamma_X^\mu v(\pB_t) \bigr)
  \nonumber\\
  =&
  \frac{e^2}{s}
  \Bigl[
  \quad{}
    [v^e v^t]\,
    \bigl( \vB(\pB_e) \gamma_\mu u(p_e) \bigr)
    \bigl( \uB(p_t) \gamma^\mu v(\pB_t) \bigr)
  \nonumber\\
  &\qquad{}
    -
    [v^e a^t]\,
    \bigl( \vB(\pB_e) \gamma_\mu u(p_e) \bigr)
    \bigl( \uB(p_t) \gamma^\mu\gamma_5 v(\pB_t) \bigr)
  \nonumber\\
  &\qquad{}
    -
    [a^e v^t]\,
    \bigl( \vB(\pB_e) \gamma_\mu\gamma_5 u(p_e) \bigr)
    \bigl( \uB(p_t) \gamma^\mu v(\pB_t) \bigr)
  \nonumber\\
  &\qquad{}
    +
    [a^e a^t]\,
    \bigl( \vB(\pB_e) \gamma_\mu\gamma_5 u(p_e) \bigr)
    \bigl( \uB(p_t) \gamma^\mu\gamma_5 v(\pB_t) \bigr)
  \Bigr]
  \sepcomma
\end{align}
where
\begin{align}
  &  [v^e v^t] =
     v^{e\gamma}v^{t\gamma} + d(s)\,v^{eZ}v^{tZ}
  \sepcomma\nonumber\\
  &  [v^e a^t] =
     v^{e\gamma}a^{t\gamma} + d(s)\,v^{eZ}a^{tZ}
     = d(s)\,v^{eZ}a^{tZ}
  \sepcomma\nonumber\\
  &  [a^e v^t] =
     a^{e\gamma}v^{t\gamma} + d(s)\,a^{eZ}v^{tZ}
     = d(s)\,a^{eZ}v^{tZ}
  \label{eq:eff-coupling}
  \sepcomma\\
  &  [a^e a^t] =
     a^{e\gamma}a^{t\gamma} + d(s)\,a^{eZ}a^{tZ}
     = d(s)\,a^{eZ}a^{tZ}
  \nonumber
\end{align}
represent energy-dependent ``couplings''.
Extensions to the anomalous vertices should be obvious.  
$d(s)$ is the ratio of the $Z$ propagator to the $\gamma$ propagator
\begin{align}
  d(s) \equiv \frac{\gZ^2}{e^2}\frac{s}{s-m_Z^2+im_Z\Gamma_Z}
  \label{eq:ds-def=prop-ratio}
\end{align}
with
\begin{align}
  \frac{\gZ^2}{e^2} 
  = \frac{\ds\pfrac{g}{\cos\thetaW}^2}{(g\sin\thetaW)^2}
  = \frac{1}{\cos^2\thetaW\,\sin^2\thetaW} 
  = 5.625
  \sepperiod
\end{align}
The width $\Gamma_Z$ of $Z$ introduces an absorptive part.
At $\sqrt{s} = 2 \times 175\GeV$, its relative magnitude is 
$s/(s-m_Z^2+im_Z\Gamma_Z) = 1.073 - 0.002\,i$.  
Thus, in the threshold region, 
the Coulomb binding effects overwhelm the effect of $\Gamma_Z$.  
Also in the open-top region, it is known that 
the QCD correction is larger than the effect of $\Gamma_Z$, 
as far as the normal component of the polarization of 
top quark $\Polnorm$ is concerned.

\end{appendix}

\newpage

\def\app#1#2#3{{\it Acta~Phys.~Polonica~}{\bf B #1} (#2) #3}
\def\apa#1#2#3{{\it Acta Physica Austriaca~}{\bf#1} (#2) #3}
\def\npb#1#2#3{{\it Nucl.~Phys.~}{\bf B #1} (#2) #3}
\def\plb#1#2#3{{\it Phys.~Lett.~}{\bf B #1} (#2) #3}
\def\prd#1#2#3{{\it Phys.~Rev.~}{\bf D #1} (#2) #3}
\def\pR#1#2#3{{\it Phys.~Rev.~}{\bf #1} (#2) #3}
\def\prl#1#2#3{{\it Phys.~Rev.~Lett.~}{\bf #1} (#2) #3}
\def\sovnp#1#2#3{{\it Sov.~J.~Nucl.~Phys.~}{\bf #1} (#2) #3}
\def\yadfiz#1#2#3{{\it Yad.~Fiz.~}{\bf #1} (#2) #3}
\def\jetp#1#2#3{{\it JETP~Lett.~}{\bf #1} (#2) #3}
\def\zpc#1#2#3{{\it Z.~Phys.~}{\bf C #1} (#2) #3}

\newpage

\end{document}